\documentclass{!!Arti}
\usepackage[leqno]{amsmath}
\usepackage[psamsfonts]{amssymb}
\usepackage{amsthm}
\usepackage[mathscr,psamsfonts]{eucal}
\usepackage{cmmib57}
\usepackage{!Style}
\usepackage{!Macros}
\usepackage{amscd}
\usepackage{url}
\usepackage[dvips,%
pdftitle={A covariant approach to the quantisation of a rigid body},%
pdfauthor={M. Modugno, C. Tejero Prieto, R. Vitolo},%
pdfkeywords={Covariant classical mechanics, covariant quantum
  mechanics, rigid body.}%
]{hyperref}
\usepackage[midshaft,scriptlabels]{diagrams}
\newcommand{\pat}{}%{_{\text{pat}}}
\newcommand{\cen}{_{\text{cen}}}
\newcommand{\dia}{_{\text{dia}}}
\newcommand{\rel}{_{\text{rel}}}
\newcommand{\rig}{_{\text{rig}}}
\newcommand{\mul}{_{\text{mul}}}
\newcommand{\anl}{_{\text{ang}}}
\newcommand{\rot}{_{\text{rot}}}

\DeclareMathOperator{\op}{Op}
\title{\bf A covariant approach to
\\
the quantisation of a rigid body}
\author{
\bf Marco Modugno$^1$, Carlos Tejero Prieto$^2$, Raffaele Vitolo$^3$
\bigskip
\\
\fz $^1$Department of Applied Mathematics ``G. Sansone",
University of Florence
\\
\fz Via S. Marta 3, 50134 Florence, Italy
\\
\fz email: {\tt marco.modugno@unifi.it}
\myskip
\\
\fz $^2$Department of Mathematics, University of Salamanca
\\
\fz Plaza de la Merced 1-4, 37008 Salamanca, Spain
\\
\fz email: {\tt carlost@usal.es} 
\myskip
\\
\fz $^3$Department of Mathematics ``E. De Giorgi", University of Lecce
\\
\fz Via per Arnesano, 73100 Lecce, Italy
\\
\fz email: {\tt raffaele.vitolo@unile.it}
\vspace{-2cm}}
\edition
{\fz {\em Preprint 2005.11.01. - 08.30.}}
\date{}
\begin{document}
%--------------------------------------------------------------------%
\maketitle
%--------------------------------------------------------------------%
\begin{abstract}
This paper concerns the quantisation of a rigid body in the
framework of ``covariant quantum mechanics'' on a curved spacetime
with absolute time.

The basic idea is to consider the multi-configuration space,
i.e. the configuration space for $n$ particles, as the $n$-fold
product of the configuration space for one particle.  Then we impose a
rigid constraint on the multi-configuration space. The resulting space
is then dealt with as a configuration space of a single abstract
`particle'. The same idea is applied to all geometric and dynamical
structures.

We show that the above configuration space fits into the general
framework of ``covariant quantum mechanics''. Hence,  the methods
of this theory can be applied to the rigid body.

Accordingly, we find exactly two inequivalent choices of quantum
structures for the rigid body. Then, we evaluate the quantum energy
and momentum operators and the `rotational part' of their spectra. We
provide a new mathematical interpretation of two-valued wavefunctions
on $SO(3)$ in terms of single-valued sections of a new non-trivial
quantum bundle. These results have clear analogies with spin.
\end{abstract}

{\small


\myskip

{\bf Acknowledgments}

We would like to thank G. Besson, S. Gallot, A. Lopez Almorox, J.
Marsden, G. Modugno, Mi. Modugno, L. M. Navas and D. Perrone, for
useful discussions.

This work has been partially supported by MIUR, Progetto PRIN 2003
``Sistemi integrabili, teorie classiche e quantistiche'', GNFM and
GNSAGA of INdAM, and the Universities of Florence, Lecce and
Salamanca.
}
%--------------------------------------------------------------------%
\newpage
\tableofcontents
%--------------------------------------------------------------------%
\newpage
\myintro
\label{Introduction}
%--------------------------------------------------------------------%
A covariant formulation of classical and quantum mechanics on a
curved spacetime with absolute time based on fibred manifolds,
jets, non linear connections, cosymplectic forms and Fr\"olicher
smooth spaces has been proposed by A. Jadczyk and M. Modugno
\cite{JadMod92,JadMod94} 
and further developed by several authors
(see, for instance, 
\cite{CanJadMod95,
%JadJanMod98,Jan95a,Jan95b,Jan98,Jan01,
%JanMod96a,JanMod96b,JanMod97,JanMod98,
%JanMod99,JanMod00,JanMod01,
JanMod02a,
%JanMod02b,
JanMod02c,
%JanMod05p1,
JanModSal02,
%JanMod05p1,Mod98,ModVit96,
SalVit00,
%Vit96,
Vit99,ModTejVit00}). 
We shall briefly call this approach ``covariant quantum mechanics''.
It presents analogies with geometric quantisation (see, for instance,
\cite{Gar79,Got98,Kos70,Sou70,Sni80,Woo92} and references
therein), but several novelties as well. In fact, it overcomes
typical difficulties of geometric quantisation such as the problem
of polarisations; moreover, in the flat case, it reproduces the
standard quantum mechanics, hence it allows us to recover all
classical examples
(see \cite{ModTejVit00} for a comparison between the two approaches).

Here, we discuss an original geometric formulation of classical and
quantum mechanics for a rigid body according to the general scheme of
``covariant quantum mechanics''. Our method, based on the classical
multi--body and rigid model developed in detail in
\cite{ModVit05p} and on the covariant quantum mechanics, seems to be a
new approach, which is able to unify different cases on a clean
mathematical scheme.

\myskip

We start with a sketch of the essential features of the general
``covariant quantum mechanics'' following
\cite{JadMod94,JanMod02c,JanMod05p1,Vit99}. The classical theory is
based on a fibred manifold (``spacetime'') over time, equipped with a
vertical Riemannian metric (``space--like metric''), a certain
time and metric preserving linear connection (``gravitational
connection'') and a closed 2--form (``electromagnetic field'').
The above objects yield a cosymplectic 2--form on the first jet
space of spacetime (``phase space''), in the sense of
\cite{deLTuy96}. This 2--form controls the classical dynamics. The
quantum theory is based on a Hermitian line bundle over spacetime
(``quantum bundle'') equipped with a Hermitian universal
connection, whose curvature is proportional to the above classical
cosymplectic $2$-form. This quantum structure yields in a natural
way a Lagrangian (hence the dynamics) and the quantum operators.

\myskip

In view of the formulation of classical mechanics of a rigid body in
the framework of the above scheme, we proceed in three steps
\cite{ModVit05p}.

Namely, we start with a flat spacetime for a pattern one--body
mechanics.

Then, we consider an $n$-fold fibred product of the pattern
structure as multi-spacetime for the $n$-body mechanics. A
geometric `product space' for $n$-body mechanics has been
developed by several authors in different ways (see, for instance,
\cite{CurMil85,DoeMan98,Mar96,JadMod94,ModVit05p}. In particular,
our approach is close to that in \cite{CurMil85} for the
`rotational' part of the rigid body dynamics, it can be easily
compared with \cite{MarRat94}, and is close to \cite{DoeMan98} for
the formulation of quantum structures.

Moreover, we consider the subbundle of the multi--spacetime
induced by a rigid constraint as configuration space for the rigid
body mechanics. We can prove that this configuration space
fulfills the requirements of the spacetime assumed in ``covariant
quantum mechanics''; hence, the general machinery of covariant
classical and quantum mechanics can be easily applied to the rigid
body.

Next, we proceed with the quantisation of the rigid body.
We discuss the existence and classification of the inequivalent
quantum structures over the rigid configuration space. 
Quantum structures are pairs consisting of a hermitian complex line
bundle and a hermitian connection on it whose curvature is
proportional to the cosymplectic
$2$-form.  
It turns out that there are two possible quantum bundles: a trivial
and a non trivial one.  
The transition functions of the non-trivial bundle are constant, hence
both the trivial and the non-trivial bundle are endowed with a flat
hermitian connection. Such connections can be deformed by adding the
classical Poincar\'e--Cartan form to produce two non-isomorphic
quantum structures.

Then, we evaluate the classical `translational' and `rotational'
observables of position, momenta ad energy and the corresponding
quantum operators.

Finally, we explicitly compute the spectra of the rotational momentum
and energy quantum operators for all quantum structures, in the case
of vanishing electromagnetic field (`free' rigid body). The
computations existing in the literature for the spectra of a rigid
body in some special electromagnetic fields can be recovered in our
scheme analogously to the previous procedure by means of the quantum
structure associated with the trivial quantum bundle (see, for
instance, \cite{Are70,BarMikShi01,BarBozMar92,BarDur91,BozArs94,Cas31,
ChoSmi65,CohDiuLal77,DuvElhGotTuy90,Got64,HajOpa91,Her44,Hun77,Iwa99,
Joh80,Kom01,LanLif67,LitRei97,LitMit01,Mar03,Mes59,PanDra99,Pav04,
Put93,SlaSla91,TanIwa99,Tej01,Tot97,Tru97}).

The non-trivial quantum structure is an original feature of our paper
in the framework of covariant quantum mechanics; for a similar result
from the viewpoint of geometric quantisation, see
\cite{Tej01}. The non-trivial quantum bundle provides a clear
mathematical setting and interpretation of the double--valued
wavefunctions formalism. Indeed, several authors have consider
double--valued wavefunctions (see, for instance,
\cite{BarBozMar92,BozArs94,CohDiuLal77,
Got64,LanLif67,Mes59,PanDra99}). Sometimes \cite{CohDiuLal77,Mes59},
these functions have been discarded because they are supposed to
break the continuity of the quantum rotation operator. However, in
our approach, no continuity is broken if we allow the existence of a
non trivial quantum bundle.  Schr\"odinger refused to consider
double-valued wavefunctions. But by some authors \cite{KliKon84} is
was argued that the probability density must be single valued, hence
double-valued wavefunctions must be accepted because their square is
single-valued.

However, the most important contribution on this problem was given by
Casimir in his Ph.D. thesis \cite{Cas31}.
On p.\ 72 Casimir explains that the two-valuedness is due to the
non-contractibility of the space of rotations:
\begin{quotation}
  \dots To a curve connecting
$\xi \,,$
$\eta \,,$
$\zeta \,,$
$\chi$
with
$-\xi \,,$
$-\eta \,,$
$-\zeta \,,$
$-\chi$
there corresponds a closed motion
  that cannot be contracted; it may be changed into a rotation through
$2\pi \,.$
Accordingly, we may say: the two-valuedness of the
$\xi_i$
[coordinates on
$\B R^4$
restricted to
$S^3$
], \dots, 
the possibility of two-valued representations, are based on the
kinematical fact that a
$2\pi$
rotation cannot be contracted.
\end{quotation}
In our opinion, our non-trivial quantum bundle is a modern
topological model implementing the above features: it appears exactly
because the fundamental group of
$SO(3)$
is
$\B Z_2 \,.$

As far as we know, in all cases in which the spectrum of molecules
has half--integer eigenvalues, this value can be attributed to the
spin of constituent particles. It means that nature chooses
always, by a superselection rule, the trivial bundle. On the other
hand, our scheme foresees the non trivial bundle as another
theoretical possibility. Even if this one seems to have no actual
physical reality, it could be taken as basis for a kind of
``semi--classical model of spin'', by taking into account the
scheme of covariant quantum mechanics for a spin particle
\cite{CanJadMod95}. Indeed, some authors have considered such a
possibility, by following other approaches (see, for instance
\cite{BarBozMar92,BozArs94,Got64,HanReg74}).

% beginning of the part added by raf 2005.11.05
\myskip

In this paper we compute spectra only with respect to a fixed inertial
observer. However, as a by-product of the covariant approach to
quantum rigid systems, we could compute spectra also with respect to
accelerated observers. In experiments it often happens that spectra
come from sources which do not have inertial motion with respect to
the laboratory. It is customary to add ad hoc terms to the standard
Schr\"odinger operator in order to fit most spectral lines. Our
framework allows one to obtain covariant Schroedinger operators with
respect to accelerated frames. Hence, in principle, it should be
possible to compute explicitly (possibly by means of numerical
analysis techniques) the spectra. But this issue is left to future
investigations.

\myskip
%end of the part added by raf 2005.11.05

In this paper we will not touch the issue of reduction. In
particular, it would be interesting to check if the
Guillemin--Sternberg conjecture \cite{GuiSte82} (see also
\cite{Got86}) holds in the case of a free rigid body.  (We
recall that the Guillemin--Sternberg conjecture states the
commutation between the reduction and the quantization
procedures.) In fact, the group $SO(3)$ acts as a group of
symmetries on a free rigid body. A cosymplectic reduction
procedure (analogous to the Marsden--Weinstein reduction
procedure) could be formulated. A similar analysis has been
carried out in \cite{Joh80} in order to formulate a geometric
prequantization (see also \cite{Put93} for similar results under
stronger hypotheses). The coadjoint orbits of constant angular
momentum turn out to be spheres $S^2 \,.$ It would be very
interesting to investigate the interplay between the two
inequivalent quantum structures of the rigid body and the possible
quantum structures of coadjoint orbits, especially in view of the
fact that their topologies are different. But this will be the
subject of future work.

\myskip

From a physical viewpoint, our model can describe extremely cold
molecules. In fact, vibrational modes are of great importance in
quantum dynamics, unless the temperature is extremely low. A
different approach to the quantisation of a rigid body is provided
by the so called `pseudo-rigid body'
\cite{DuvElhGotTuy90,Mar96,Tru97}, by considering a potential with
suitable wells confining bodies to be near to a rigid constraint.
This approach seems to be more physical than ours, but it is more
complicated. Indeed, we think that our approach can be considered
a useful model, due to its simplicity. Even in the purely
classical description of rigid bodies one can follow two ways: i)
a more physical but very complicated one, by considering forces
bounding the constituent particles and by referring to the limit
case when these forces freeze the distances between the particles;
ii) a more ideal and much simpler one, by considering the rigid
constrained body, regardless of the physical origin of the
constraint. The scheme of covariant quantum mechanics allows us to
apply a viewpoint analogous to the second approach mentioned above to
the quantum mechanics of a ``rigid body''.

\myskip

We assume manifolds and maps to be
$\Cin \,.$
If
$\f M$
and
$\f N$
are manifolds, then the sheaf of local smooth maps
$\f M \to \f N$
is denoted by
$\map(\f M, \, \f N) \,.$
%--------------------------------------------------------------------%
\newpage
\section{Covariant quantum mechanics}
\label{Covariant quantum mechanics}
%--------------------------------------------------------------------%
\setcounter{assump}{7}
\renewcommand{\theassump}{\Alph{assump}}
\setcounter{Assumption}{0}
%--------------------------------------------------------------------%
\bsm
We start with a brief sketch of the basic notions of ``covariant
quantum mechanics", paying attention just to the facts that are
strictly needed in the present paper.
We follow 
\cite{JadMod94,JanMod02c,JanMod05p1,JanMod05p2,JanModVit05,SalVit00,
Vit99}.
For further details and discussions the reader should refer to the
above literature and references therein.
\esm

In order to make classical and quantum mechanics explicitly
independent from scales, we introduce the ``spaces of scales''.
Roughly speaking, a space of scales has the algebraic structure of
$\Rn^+$ but has no distinguished `basis'. The basic objects of our
theory (metric, electromagnetic field, etc.) will be valued into {\em
scaled\/} vector bundles, that is into vector bundles twisted with
spaces of scales. We shall use rational tensor powers of spaces of
scales. In this way, each tensor field carries explicit information
on its ``scale dimension".

Actually, we assume the following basic spaces of scales:
the space of {\em time intervals\/}
$\B T \,,$
the space of {\em lengths}
$\B L \,,$
the space of {\em masses}
$\B M \,.$

We assume the {\em Planck's constant\/}
$\h \in \B T^* \ten \B L^2 \ten \B M \,.$
Moreover, a {\em particle\/} will be assumed to have a {\em mass\/}
$m \in \B M$
and a {\em charge\/}
$q \in \B T^* \ten \B L^{3/2} \ten \B M^{1/2} \,.$
%--------------------------------------------------------------------%
\subsection{Classical scheme}
\label{Classical scheme}
%--------------------------------------------------------------------%
\bAs
We assume:

- the {\em time\/} to be an affine space
$\f T$
associated with the vector space
$\baB T \byd \B T \ten \Rn \,,$

- the {\em spacetime\/} to be an oriented manifold
$\f E$
of dimension
$1+3 \,,$

- the {\em time fibring\/} to be a fibring (i.e., a surjective
submersion)
$t : \f E \to \f T \,,$

- the {\em spacelike metric\/} to be a scaled vertical Riemannian
metric
\beq
g : \f E \to \B L^2 \ten S^2 V^*\f E \,,
\eeq

- the {\em gravitational connection\/} to be a linear connection
of spacetime 
\beq 
K\Nat : T\f E \to T^*\f E \uten{T\f E} TT\f E \,, 
\eeq 
such that 
$\nab[K\Nat] dt = 0$ 
and 
$\nab[K\Nat] g = 0 \,,$ 
and whose curvature 
$R[K\Nat]$ 
is ``vertically symmetric",

- the {\em electromagnetic field\/} to be a closed scaled 2--form
\beq
F : \f E \to (\B L^{1/2} \ten \B M^{1/2}) \ten \Lam^2T^*\f E \,.
\eeq

The spacelike orientation and the metric
$g$
yield the spacelike scaled volume form
$\eta$
and
its dual
$\ba\eta \,.$

With reference to a given particle with mass
$m$
and charge
$q \,,$
it is convenient to consider the {\em rescaled\/} sections
\beq
G \byd \tfr m\h g : \f E \to \B T \ten S^2 V^*\f E
\ssep{and}
\tfr q\h F : \f E \to \Lam^2T^*\f E \,.
\eeq

We shall refer to fibred charts
$(x^{0}, x^i)$
of
$\f E \,,$
where
$x^0$
is adapted to the affine structure of
$\f T$
and to a time scale
$u_0 \in \B T \,.$
Latin indices
$i,j, \dots$
and Greek indices
$\lam, \mu, \dots$
will label space--like and spacetime coordinates, respectively.
For short, we shall denote the induced dual bases of vector fields and
forms by
$\der_\lam$
and
$d^\lam \,.$
The vertical restriction of forms will be denoted by the check
$``\,^\vee\," \,.$

We have the coordinate expression
$G = G^0_{ij} \, u_0 \ten \ch d^i \ten \ch d^j \,.$
The coordinate expression of the condition of vertical symmetry of
$R[K\Nat]$ is
$R\Nat_{i\lam j\mu} =
R\Nat_{j\mu i\lam} \,.$

\myskip

A {\em motion\/} is defined to be a section
$s : \f T \to \f E \,.$

We assume the first jet space of motions
$J_1\f E$
as {\em phase space\/} for classical
mechanics of a spinless particle; the first jet prolongation
$j_1s$
of a motion $s$ is said to be  its {\em velocity\/}. We denote by
$(x^\lam, x^i_0)$
the chart induced on
$J_1\f E \,.$
We shall use the natural  complementary maps
$\K d : J_1\f E \car \baB T \to T\f E$
and
$\tht : J_1\f E \car_{\f E} T\f E \to V\f E \,,$
with coordinate expressions
$\K d = u^0 \ten (\der_0 +  x^i_0 \, \der_i)$
and
$\tht = (d^i - x^i_0 \, d^0) \ten \der_i \,.$
We set
$\tht^i \eqv d^i - x^i_0 \, d^0 \,.$

An {\em observer\/} is defined to be a (local) section
$o : \f E \to J_1\f E \,.$

An observer $o$ is said to be {\em rigid\/} if the Lie derivative
$L[o] \, g$
vanishes.

\myskip

Let us consider an observer
$o \,.$

A chart
$(x^0, x^i)$
is said to be {\em adapted\/} to $o$ if
$o^i_0 \eqv x^i_0 \com o = 0 \,.$
We obtain the maps
$\nu[o] : T\f E \to V\f E : X \to X - o \con dt (X)$
and
$\nab[o] : J_1\f E \to \B T^* \ten V\f E : e_1 - o(e) \,.$
We define the {\em observed component\/} of a vector
$v \in T\f E \,,$
to be the spacelike vector
$\ve v[o] \byd \nu[o] (v) \,.$
Accordingly, if
$s$ is a motion, then we define the {\em observed velocity\/} to be
the section
$\nab[o] \com j_1s : \f T \to \B T^* \ten V\f E \,.$

We define the {\em observed kinetic energy\/} and {\em momentum\/},
respectively, as the maps
\beq
\C K[o] \byd \tfr12 \, G (\nab[o], \nab[o]) : J_1\f E \to T^*\f E
\sep{and}
\C Q[o] \byd \tht^* \com G\Fla (\nab[o]) : J_1\f E \to T^*\f E \,,
\eeq
with coordinate expressions
$\C K[o] = \Kin ij \, d^0$
and
$\C Q[o] = \Mom ij \, \tht^i \,.$

We define the {\em magnetic field\/} and the {\em observed electric
field\/}, respectively, as
\beq
\ve B \byd \tfr12 i(\wch F) \ba\eta
\ssep{and}
\ve E[o] \byd - o \con F \,.
\eeq
Then, we obtain the observed splitting
\beq
F = - 2 dt \wed \ve E[o] + 2 \nu^*[o] \big(i(\ve B) \eta\big) \,.
\eeq

\myskip

The linear connection
$K\Nat$
yields
an affine connection
$\Gam\Nat$
of the affine bundle
$J_1\f E \to \f E \,,$
with coordinate expression
$\Gam\Nat\Gaa \lam i \mu = K\Nat\col \lam i \mu \,,$
and the non linear connection
$\gam\Nat \byd \K d \con \Gam\Nat : J_1\f E \to \B T^* \ten TJ_1\f E$
of the fibred manifold
$J_1\f E \to \f T \,,$
with coordinate expression
$\gam \Nat =
u^0\ten (\der_0 + x^i_0 \, \der_i + \gam\Nat\ga i \, \der_i^0) \,,$
where
$\gam\Nat\ga i \byd
K\Nat\col hik \, x^h_0 \, x^k_0 + 2 \, K\Nat\col hi0 \, x^h_0 +
K\Nat\col 0i0 \,.$
Moreover,
$\Gam\Nat$
yields the $2$--form
$\Ome\Nat \byd G(\nu[\Gam\Nat] \wed \tht) :
J_1\f E \to \Lam^2 T^*J_1\f E \,.$
We have the coordinate expression
$\Ome\Nat =
G^0_{ij} \,
(d^i_0 - \gam\Nat\ga i \, d^0
- \Gam\Nat\Ga hi \, \tht^h) \wed \tht^j \,,$
where
$\Gam\Nat\Ga hi \eqv
\Gam\Nat\Gaa hik \, x^k_0 + \Gam\Nat\Gaa hi0 \,.$
The
$2$--form
$\Ome\Nat$
turns out to be closed, in virtue of the assumed symmetry of
$R[K\Nat] \,,$
and {\em non degenerate\/} as
$dt \wed \Ome\Nat \wed \Ome\Nat \wed \Ome\Nat$
is a scaled volume form of
$J_1\f E \,.$
Thus,
$\Ome\Nat$
turns out to be a {\em cosymplectic form\/}.

\myskip

There is a natural geometric way to
``merge" the gravitational and electromagnetic objects into {\em
joined\/} objects, in such a way that all mutual relations holding for
gravitational objects are preserved for joined objects.
Later on, we shall refer to such joined objects and we can forget
about the two component fields, in many respects.
In particular, we deal with the joined $2$--form
$\Ome \byd \Ome\Nat + \tfr12 \fr q\h \, F$
and the joined connection
$\gam = \gam\Nat + \gam^e \,,$
where
$\gam^e$
turns out to be the Lorentz force
$\gam^e = - G\Sha \oset{\vee}{(\K d \con F)} \,.$

We obtain
$d \Ome = 0$
and
$dt \wed \Ome \wed \Ome \wed \Ome =
dt \wed \Ome\Nat \wed \Ome\Nat \wed \Ome\Nat \,.$
Thus, also
$\Ome$
turns out to be a {\em cosymplectic form\/}.

The joined 2--form
$\Ome$
rules the classical dynamics in the following way.

The closed form
$\Ome$
admits local ``horizontal" potentials of the type
$A\Upa : J_1\f E \to T^*\f E \,,$
whose coordinate expression is of the type
$A\Upa = - (\Kin ij - A_0) \, d^0 + (\Mom ij + A_i) \, d^i \,,$
that is, for each observer
$o \,,$
of the type
$A\Upa = - \C K[o] + \C Q[o] + A[o] \,,$
where
$A[o] \byd o^* A\Upa : \f E \to T^*\f E \,.$

We define the (local) {\em Lagrangian\/}
$\C L[A\Upa] \byd \K d \con A\Upa : J_1\f E \to T^*\f E$
and the (local) {\em momentum}
$\C P[A\Upa] \byd \tht^* V_{\f E}\C L[A\Upa] :
J_1\f E \to T^*\f E \,,$
with expressions
$\C L[A\Upa] = (\Kin ij + A_i \, x^i_0 + A_0) \, d^0$
and
$\C P[A\Upa] = (\Mom ij + A_i) \, \tht^i \,.$
Indeed,
the Poincar\'e--Cartan form
$\The = \C L[A\Upa] + \C P[A\Upa]$
associated with
$\C L[A\Upa]$
turns out to be just
$A\Upa \,.$

Moreover, given an observer
$o \,,$
we define the (observed) {\em Hamiltonian\/}
$\C H[A\Upa, o] \byd\linebreak - o \con A\Upa : J_1\f E \to T^*\f E$
and the (observed) {\em momentum\/}
$\C P[A\Upa, o] \byd \nu[o] \con A\Upa : J_1\f E \to T^*\f E \,,$
with coordinate  expressions
$\C H[A\Upa, o] = (\Kin ij - A_0) \, d^0$
and
$\C P[A\Upa, o] = (\Mom ij + A_i) \, d^i \,,$
in adapted coordinates.
We obtain also the scaled function
$\|\C P[A\Upa, o]\|^2 \,,$
with coordinate expression
$\|\C P[A\Upa, o]\|^2_0 =
G{_{ij}^0} \, x^i_0 \, x^j_0 + 2 \, A_i \, x^i_0 +
G{^{ij}_0} \, A_i \, A_j\,.$

The Euler--Lagrange equation, in the unknown  motion
$s \,,$
associated with the (local) Lagrangians
$\C L[A\Upa]$
turns out to be the global equation
$\nab[\gam] j_1s = 0 \,,$
that is
$\nab[\gam\Nat] j_1s = \gam^e \com j_1s \,.$
This equation is just the generalised Newton's equation of motion for
a charged particle in the given gravitational and electromagnetic
field.
We assume this equation to be our classical equation of motion.
%--------------------------------------------------------------------%
\subsection{Quantum scheme}
\label{Quantum scheme}
%--------------------------------------------------------------------%
A {\em quantum bundle\/} is defined to be a complex line bundle
$\f Q \to \f E \,,$
equipped with a Hermitian metric
$\E h$
with values in
$\B C \ten \Lam^3 V^*\f E \,.$
A {\em quantum section\/}
$\Psi :\f E \to \f Q$
describes a quantum particle.

A local section
$\E b : \f E \to \B L^{3/2} \ten \f Q \,,$
such that
$\E h(\E b, \E b) = \eta \,,$
is a local basis.
We denote the local complex dual basis of
$\E b$
by
$z : \f Q \to \B L^{*3/2} \ten \B C \,.$
If
$\Psi$
is a quantum section, then we write locally
$\Psi = \psi \E b \,,$
where
$\psi \byd z \com \Psi : \f E \to \B L^{*3/2} \ten \B C \,.$

The {\em Liouville vector field\/} is defined to be the vector field
$\B I : \f Q \to V\f Q : q \mto (q,q) \,.$

Lets us consider the {\em phase quantum bundle\/}
$\f Q\Upa \byd J_1\f E \car_{\f E} \f Q \to J_1\f E \,.$

If
$\{\K Q[o]\}$
is a family of Hermitian connections of
$\f Q \to \f E$
parametrised by the observers
$o \,,$
then there is a unique Hermitian connection
$\K Q\Upa$
of
$\f Q\Upa \to J_1\f E \,,$
such that
$\K Q[o] = o^* \K Q\Upa \,,$
for each observer
$o \,.$
This connection is called {\em universal\/} and is locally of the type
$\K Q\Upa = \chi\Upa[\E b] + \coi A\Upa[\E b] \ten \B I\Upa \,,$
where
$\chi\Upa[\E b]$
is the flat connection induced by a local quantum basis
$\E b$ and $A\Upa[\E b]$
is a local horizontal $1$--form of
$J_1\f E \,.$
The map
$\{\K Q[o]\} \mto \K Q\Upa$
is a bijection.

We define a {\em phase quantum connection\/} to be a connection
$\K Q\Upa$
of the phase quantum bundle, which is Hermitian, universal and whose
curvature is
$R[{\K Q\Upa}] = - 2 \, \coi \Ome \ten \B I\Upa \,.$

A phase quantum connection 
$\K Q\Upa$ 
is locally of the type
$\K Q\Upa = \chi\Upa[\E b] + \coi A\Upa[\E b] \ten \B I\Upa \,,$ 
where 
$A\Upa[\E b]$ 
is a local horizontal potential for 
$\Ome \,.$

We remark that the equation
$d\Ome = 0$
turns out to be just the Bianchi identity for a phase quantum
connection
$\K Q\Upa \,.$

A pair 
$(\f Q, \K Q\Upa)$ 
is said to be a {\em quantum structure}.

Two quantum bundles
$\f Q_1$
and
$\f Q_2$
on
$\f E$
are said to be {\em equivalent\/} if there exists an isomorphism of
Hermitian line bundles
$f : \f Q_1 \to \f Q_2$
over
$\f E$
(the existence of such an
$f$
is equivalent to the existence of an isomorphism of line bundles).
Two quantum structures
$(\f Q_1 ,\K Q\Upa_1)$
and
$(\f Q_2 ,\K Q\Upa_2) \,,$
are said to be {\em equivalent\/} if there exists an equivalence
$f : \f Q_1 \to \f Q_2$
which maps
$\K Q\Upa_1$
into
$\K Q\Upa_2 \,.$

A quantum bundle is said to be {\em admissible\/} if it admits a
phase quantum connection.
Actually, the following results holds.

Let us consider the cohomology
$H^*(\f E, X)$
with values in
$X = \Rn \,,$
or
$X = \B Z \,,$
the inclusion morphism
$i : \B Z \to \Rn$
and  the induced group morphism
$i^* : H^*(\f E, \B Z) \to H^*(\f E, \Rn) \,.$

The difference of two local horizontal potentials for $\Ome$ turns
out to be a locally closed spacetime form. Therefore, we can prove
that the de Rham class 
$[\Ome]_R$ 
naturally yields  a cohomology class 
$[\Ome] \in H^2(\f E, \Rn) \,.$

\bPr\label{existence and classification} \cite{Vit99} We
have the following classification results.

1) The equivalence classes of complex line bundles on
$\f E$
are in bijection with the 2nd cohomology group
$H^2(\f E, \B Z) \,.$

2) There exists a quantum structure on
$\f E \,,$
if and only if
\beq
[\Ome] \in
i^2 \big(H^2(\f E, \B Z)\big) \sub H^2(\f E, \Rn) \seq
H^2(J_1\f E, \Rn) \,.
\eeq

3) Equivalence classes of quantum structures are in bijection with
the  set
\beq
(i^2)^{-1}([\Ome]) \; \car \;
H^1(\f E, \Rn) \big/ H^1(\f E, \B Z ) \,.
\eeq

More precisely, the first factor parametrises admissible quantum
bundles and the second factor parametrises phase quantum
connections.\END
\ePr

The quantum theory is based on the only assumption of a quantum
structure, supposing that the background spacetime admits one.

\bAs
We assume a quantum bundle
$\f Q$
equipped with a phase quantum connection
$\K Q\Upa \,.$\END
\eAs

All further quantum objects will be derived from the above quantum
structure by natural procedures.

We have been forced to assume that
$\K Q\Upa$
lives on the phase quantum bundle
$\f Q\Upa$
because of the required link with the
$2$--form $\Ome \,.$
On the other hand, in order to accomplish the covariance of the
theory, we wish to derive from
$\K Q\Upa$
new quantum objects, which are observer independent, hence living on
the quantum bundle. For this purpose we follow a successful
projectability procedure: if
$\f V\Upa \to J_1\f E$
is a vector bundle which projects on a vector bundle
$\f V \to \f E \,,$
then we look for sections
$\sig\Upa: J_1\f E \to \f V\Upa$
which are projectable on sections
$\sig: \f E \to \f V$
and take these
$\sig$
as candidates to represent quantum objects.

\myskip

The quantum connection allows us to perform covariant derivatives of
sections of
$\f Q$
(via pullback).
Then, given an observer
$o \,,$
the {\em observed quantum connection\/}\linebreak
$\K Q[o] \byd o^*\K Q\Upa$
yields, for each section
$\Psi : \f E \to \f Q \,,$
the {\em observed quantum differential\/} and the {\em observed
quantum Laplacian\/}, with coordinate expressions
\bal
\ob\nab_\lam \, \psi
&=
(\der_\lam - \coi A_\lam) \, \psi \,,
\\
\ob\Del_0 \, \psi
&=
\big(G^{hk}_0 \, (\der_h - \coi A_h) (\der_k - \coi A_k) +
\diG h k g (\der_k - \coi A_k)\big) \, \psi \,.
\end{align*}

We can prove that all 1st order covariant quantum Lagrangians
\cite{JanMod02c} are of the type (we recall that
$m/\h$ has been incorporated into $G$ and $A[o]$)
\bml
\E L[\Psi]
= \tfr12 \, \big( \coi (\ba\psi \, \der_0 \psi - \psi \, \der_0
\ba\psi) + 2 \, A_0 \, \ba\psi \, \psi
\\
- G^{ij}_0 \,
(\der_i \ba\psi \, \der_j \psi + A_i \, A_j \, \ba\psi \, \psi)
- \coi A^i_0 \, (\ba\psi \, \der_i \psi - \psi \, \der_i \ba\psi)
+
k \, \rho_0 \, \ba\psi \, \psi\big)
\;
\rtd g \, d^0 \wed d^1 \wed d^2 \wed d^3 \,,
\end{multline*}
where
$\rho_0 = G^{ij}_0 \, R\cul hihj$
is the scalar curvature of the spacetime connection
$K$
and
$k \in \Rn$
is an arbitrary parameter (which cannot be determined by covariance
arguments).

By a standard procedure, these Lagrangians yield the quantum momentum,
the Euler--Lagran\-ge operator (generalised {\em Schr\"odinger
operator\/}) and a conserved form ({\em probability current\/}).
We assume the quantum sections $\Psi$ to fulfill the generalised
Schr\"odinger equation with coordinate expression (we recall that
$m/\h$
has been incorporated into
$G$
and
$A[o]$)
\beq
\E S_0 \, \psi = \big(\ob\nab_0 + \tfr12 \, \di 0 g -
\tfr12 \, \coi \, (\ob\Del_0 + k \, \rho_0) \,
\big) \, \psi = 0 \,.
\eeq

Next, we sketch the formulation of quantum operators.

We can exhibit a distinguished Lie algebra
$\spec(J_1\f E) \sub \map(J_1\f E, \Rn)$
of functions, called {\em special phase functions\/}, of the type
$f = f^0 \, \Kin ij + f^i \, \Mom ij + \br f \,,$
where
$f^\lam, \br f \in \map(\f E, \Rn) \,.$
Among special phase functions we have
$x^\lam \,,$
$\C P_j \,,$
$\C H_0$
and
$\|\C P\|^2_0 \,.$
The bracket of this algebra is defined in terms of the Poisson
bracket and
$\gam \,.$

Then, by classifying the vector fields
on $\f Q$ which preserve the Hermitian metric and are
projectable on $\f E$ and on $\f T \,,$ we see that they
constitute a Lie algebra, which is naturally isomorphic to the Lie
algebra of special phase functions. These vector fields can be
regarded as {\em pre--quantum operators\/} $Z[f]$ acting on
quantum sections.

\myskip

The {\em sectional quantum bundle\/} is defined to be the bundle
$\ha{\f Q} \to \f T \,,$
whose fibres
$\ha{\f Q}_\tau \,,$ 
with 
$\tau \in \f T \,,$ 
are constituted by smooth quantum sections, at the time
$\tau \,,$ 
with compact support. This infinite dimensional complex vector bundle
turns out to be F--smooth in the sense of Fr\"olicher
\cite{Fro82} and inherits a pre--Hilbert structure via integration
over the fibres. A Hilbert bundle can be obtained by completion.

We can prove that the Schr\"odinger operator
$\E S$
can be naturally regarded as a linear connection of
$\ha{\f Q} \to \f T \,.$

Eventually, a natural procedure associates with every special phase
function
$f$
a symmetric {\em quantum operator\/}
$\ha f : \ha{\f Q} \to \ha{\f Q} \,,$
fibred over
$\f T \,,$
defined as a linear combination of the corresponding pre--quantum
operator
$Z[f]$
and of the operator
$f^0 \, \E S_0 \,.$
We obtain the coordinate expression (we recall that
$m/\h$
has been incorporated into
$G$
and
$A[o]$)
\beq
\ha f \psi = \big(\br f
- \coi f^h \, (\der_h - \coi A_h)
- \coi \tfr12 \, \dif h g
- \tfr12 \, f^0  (\ob\Del_0 + k \, \rho_0) \big) \, \psi \,.
\eeq

For example, we have
\bgt
\wha{x^0} \, \psi = x^0 \, \psi \,,
\qquad
\wha{x^i} \, \psi = x^i \, \psi \,,
\\
\wha{\C P_j} \, \psi =
- \coi (\der_j + \tfr12 \, \di j g) \, \psi \,,
\qquad
\wha{\C H_0} \, \psi
= \big(- \tfr12 \, (\ob\Del_0 + k \, \rho_0) - A_0\big) \, \psi
\,,
\\
\wha{\|\C P\|^2_0} \, \psi =
\big(- G{^{ij}_0} \, A_i \, A_j
- \coi \diA h g - 2 \, \coi \, A^h_0 \, \der_h
- (\ob\Del_0 + k \, \rho_0) \big) \, \psi \,.
\end{gather*}
%--------------------------------------------------------------------%
\newpage
\section{Rigid body classical mechanics}
\label{Rigid body classical mechanics}
%--------------------------------------------------------------------%
\setcounter{assump}{18}
\renewcommand{\theassump}{\Alph{assump}}
\setcounter{Assumption}{0}
%--------------------------------------------------------------------%
\bsm 
Now, we consider a rigid body and show how it can be quantised
according to the scheme of the above general theory.

The configuration space of the classical rigid body is formulated in
three steps according to \cite{ModVit05p}:

- we start with a flat ``pattern spacetime'' of dimension $1+3$
for the formulation of one--body classical and quantum mechanics;

- then, we consider the $n$--fold fibred product of the pattern
spacetime, equipped with the induced structures, as the framework
for $n$--body classical and quantum mechanics;

- finally, we consider the rigid constrained fibred submanifold of
the above $n$--fold fibred product along with the induced
structures, as the framework for classical rigid--body.

Then, we show that this configuration space fits the general
setting of ``covariant quantum mechanics'' sketched in the previous
section. Hence, that general scheme can be  applied to this specific
case.
\esm
%--------------------------------------------------------------------%
\subsection{One--body mechanics}
\label{One--body mechanics}
%--------------------------------------------------------------------%
\bsm 
Following the general scheme, we start by assuming a flat
spacetime for one--body mechanics, which is called the {\em
pattern spacetime\/}. All objects related to this pattern
spacetime are called {\em pattern objects\/} 
\esm

Let us consider a system of one particle, with mass
$m$
and charge
$q \,.$

We assume as {\em pattern spacetime\/} a (1+3)--dimensional affine
space
$\f E\pat \,,$
associated with the vector space
$\baf E\pat$
and equipped with an affine map
$t\pat : \f E\pat \to \f T$
as time map.

From the above affine structure follow some immediate
consequences.

The map
$Dt\pat : \baf E\pat \to \baB T$
yields the 3--dimensional vector subspace
$\f S\pat \byd Dt^{-1}\pat (0) \sub \baf E\pat$
and the 3--dimensional affine subspace
$\f U\pat \byd (\id[\B T^*] \ten Dt\pat)^{-1} (1)
\sub \B T^* \ten \baf E\pat \,,$
which is associated with the vector space
$\B T^* \ten \f S\pat \,.$

Thus,
$t\pat : \f E\pat \to \f T$
turns out to be a principal bundle associated with the abelian group
$\f S\pat \,.$
Moreover, we have the natural isomorphisms
$T\f E\pat \seq \f E\pat \car \baf E\pat \,,\;$
$V\f E\pat \seq \f E\pat \car \f S\pat$
and
$J_1\f E\pat \seq \f E\pat \car \f U\pat \,.$

We assume a Euclidean metric
$g\pat \in \B L^2 \ten (\f S^*\pat \ten \f S^*\pat)$
as a spacelike metric. Moreover, we assume the connection
$K\Nat\pat$
induced by the affine structure as the gravitational connection.
Furthermore, we assume an electromagnetic field
$F\pat \,.$

Thus, we obtain
$d\Ome\Nat\pat = 0$
and
$dF\pat = 0 \,.$
Moreover, because of the affine structure of spacetime,
$\Ome\Nat\pat$
and
$F\pat$
turn out to be globally exact.
We denote the global potentials (defined up to a constant) for
$\Ome\Nat\pat$
and
$F\pat$
by
$A\Upa\Nat\pat$
and
$A\Ele\pat \,.$

We recall also the obvious natural action of the group
$O(\f S\pat, g\pat)$
on
$\f S\pat \,.$

\myskip

A motion
$s$
and an observer
$o$
are said to be {\em inertial\/} if they are affine maps.
Any inertial observer yields a splitting of the type
$\f E\pat \seq \f T \car \f P\pat[o] \,,$
where
$\f P\pat[o]$
is an affine space associated with
$\f S\pat \,.$
Any inertial motion yields an inertial observer
$o$
and an affine isomorphism
$\f P\pat[o] \seq \f S\pat \,.$

For each inertial observer
$o \,,$
we obtain the splitting
$A\Upa\Nat\pat = - \C K\pat[o] + \C Q\pat[o] + A\Nat[o] \,,$
where
$A\Nat[o] \in \baf E^*$
is a constant 1--form.
%--------------------------------------------------------------------%
\subsection{Multi-body mechanics}
\label{Multi-body mechanics}
%--------------------------------------------------------------------%
\bsm
We can describe the classical mechanics of a system of $n$ particles
moving in a given gravitational and electromagnetic field by
representing this system as a one--body moving in a higher
dimensional spacetime equipped with suitable fields which fulfill the
same properties postulated for the standard spacetime.

In this way, we can use for a system of $n$ particles all concepts
and results obtained for a one--body.
\esm
%--------------------------------------------------------------------%
\mysssec{Configuration space}
\label{Multi-particle: Configuration space}
%--------------------------------------------------------------------%
\bsm
We assume as configuration space for a system of $n$ particles the
$n$--fold fibred product of the pattern spacetime, called
``multi--spacetime".
Then, the metric field, gravitational field and electromagnetic field
naturally equip this multi--spacetime with analogous ``multi" fields.

Thus, the structure of multi--spacetime is analogous to that of
pattern spacetime. The different dimension of the fibres in the
two cases has no importance in many respects; hence, most concepts
and results can be straightforwardly translated from the pattern
case to the multi--case. Indeed, the multi--fields involve
suitable weights related to the masses and charges of the
particles, in such a way that the mechanical equations arising
from the multi--approach coincide with the system of equations
for the single particles.

So, we can formulate the classical mechanics of an $n$-body
analogously to that of a one--body equipped with the total mass and
affected by the given multi--metric, multi--gravitational field and
multi--electromagnetic field.

On the other hand, the multi--spacetime is equipped with the
projections of the fibred product, which provide additional
information concerning each particle.

All objects related to this multi--spacetime are called  {\em
multi--objects\/} and labelled by the subscript
$``\mul" \,.$
\esm

Let us consider a system of
$n$
particles, with
$n \geq 2 \,,$
and with masses
$m_1, \dots, m_n$
and charges
$q_1, \dots, q_n \,.$

\myskip

Then, we define the {\em total mass\/}
$m \byd \sum_i m_i \,,$
the {\em $i$-th weight\/}
$\mu_i \byd m_i/m \in \Rn^+$
and the {\em total charge\/}
$q \byd \sum_i q_i\,.$
Of course, we have
$\sum_i \mu_i = 1 \,.$

In order to label the different particles of the system, we introduce
$n$ identical copies of the pattern objects
$\f E_i \eqv \f E\pat \,,$\;
$\f S_i \eqv \f S\pat \,,$\;
$\f U_i \eqv \f U\pat \,,$\;
$g_i \eqv g\pat \,,$\;
$F_i \eqv F\pat \,,$\;
for
$i = 1, \dots, n \,.$

We assume the fibred product over
$\f T$
\beq
\f E\mul \byd \f E_1 \ucar{\f T} \dots \ucar{\f T} \f E_n \,,
\eeq
as {\em multi--spacetime\/}, equipped with the associated
projection
$t\mul : \f E\mul \to \f T$
as {\em multi--time map\/}.

The affine multi--spacetime 
$\f E\mul$ 
is associated with the multi--vector space
$\baf E\mul = \baf E_1 \ucar{\baB T} \dots \ucar{\baB T} \baf E_n
\,,$
which turns out to be a principal bundle
$Dt\mul : \baf E\mul \to \baB T \,,$
associated with the vector space
$\f S\mul \byd \f S_1 \car \dots \car \f S_n \,.$

The group
$O(\f S\pat, g\pat)$
acts naturally component--wisely on the vector multi--space
$\f S\mul \byd \linebreak \f S_1 \car \dots \car \f S_n \,.$

Each observer
$o\pat$
yields the {\em multi--observer\/}
$o\mul \byd (o\pat \car \dots \car o\pat) \,.$

Moreover, we assume the Euclidean metrics
\beq
g\mul \byd
(\mu_1 \, g_1\car \dots \car \mu_n \, g_n)
\ssep{and}
G\mul \byd \tfr m\h g\mul \byd
(\tfr{m_1}\h \, g_1\car \dots \car \tfr {m_n}\h \, g_n)
\eeq
as {\em multi--spacelike metric\/} and {\em rescaled
multi--spacelike metric\/}, the affine connection and the 2--form
\beq
K\Nat\mul \byd
K\Nat_1 \car \dots \car K\Nat_n \ssep{and} \C F\mul \byd
(\fr{q_1}{m} F_1 \car \dots \car \fr{q_n}{m} F_n)
\quad
\eeq
as {\em multi--gravitational connection\/} and {\em rescaled
multi--electromagnetic field\/}.\END
\eAs

We define the {\em multi--magnetic field\/} and
the {\em observed multi--electric field\/}, respectively, as
\beq
\ve{\C B}\mul \byd \tfr12 i(\wchC F) \, \ba\eta\mul
\ssep{and}
\ve{\C E}\mul[o\mul] \byd - o\mul \con \C F\mul \,.
\eeq
Then, we obtain the observed splitting
\beq
\C F\mul =
- 2 \, dt\mul \wed \ve{\C E}\mul[o\mul] +
2 \, \nu^*[o\mul] \big(i(\ve{\C B}) \, \eta\mul\big) \,.
\eeq

We obtain
$d\Ome\Nat\mul = 0$
and
$d\C F\mul = 0 \,.$
Moreover,
$\Ome\mul$
and
$\C F\mul$
are globally exact.

The above multi--spacetime and multi--fields yield further several
multi--objects analogously to the case of the pattern spacetime and
pattern fields.
%--------------------------------------------------------------------%
\mysssec{Center of mass splitting}
\label{Center of mass splitting}
%--------------------------------------------------------------------%
\bsm
Due to the affine structure and the weights of masses, the
multi--spacetime is equipped with another important splitting,
which is related to the center of mass.  Namely, the
multi--spacetime splits naturally into the product of the
$3+1$--dimensional affine subspace of center of mass and the
$(3n-3)$--dimensional vector space of distances relative to the
center of mass.  This splitting will affect all geometric,
kinematical and dynamical structures, including the equation of
motion.
\esm

In view of the following definition of center of mass, let us
consider a copy
$\f E\cen \byd \f E\pat$
of the pattern spacetime, referred to as the {\em spacetime of center
of mass\/}.

We define the affine fibred projection of the {\em center of mass\/}
\beq
\pi\cen : \f E\mul \to \f E\cen :
e\mul \eqv (e_1, \dots, e_n) \mto e\cen \,,
\ssep{with}
\sum_i \,\mu_i (e_i - e\cen) \eqv 0 \,.
\eeq

We can view the space of center of mass also in another way. In
fact, let us consider the 3--dimensional diagonal affine subspace
$i\dia : \f E\dia \hto \f E\mul \,.$
Clearly, the restriction of
$\pi\cen$ to $\f E\dia$ yields an affine fibred isomorphism
$\f E\dia \to \f E\cen \,.$
We shall often identify these two spaces via the above isomorphism and
write
$i\cen : \f E\cen \hto \f E\mul \,.$

Moreover, we define the {\em center of mass space\/} and the {\em
relative space\/} to be, respectively, the 3--dimensional and the
$(3n-3)$--dimensional vector subspaces of
$\f S\mul$
\bml
\f S\cen \byd
\{v\mul \in \f S\mul \st v_1 = \dots v_n \} \,,
\qquad
\f S\rel \byd
\{v\mul \in \f S\mul \st \sum_i \mu_i \, v_i = 0\} \,.
\end{multline*}

Of course, the natural action of
$O(\f S\pat, g\pat)$
on
$\f S\mul$
restricts to a free action on
$\f S\rel \,.$

\myskip

We set
$\f E\rel \byd \f T \car \f S\rel \,.$

Then, we obtain the affine fibred splitting over
$\f T$
\bml
\f E\mul \to \f E\dia \ucar{\f T} \f E\rel =
\f E\dia \car \f S\rel \seq
\f E\cen \ucar{\f T} \f E\rel =
\f E\cen \car \f S\rel :
\\
: e\mul \mto (e\cen, \, v\rel) \byd \big(\pi\cen(e\mul), \; e\mul -
i\dia(e\cen)\big) \,.
\end{multline*}

We stress that the above splitting yields the natural projections
$\f E\mul \to \f E\dia$
and
$\f E\mul \to \f S\rel$
and the natural inclusion
$\f E\dia \to \f E\mul \,,$
but it does not yield a natural inclusion
$\f S\rel \to \f E\mul \,.$

The above splitting yields several other splittings.

\bPr
We have the following linear splittings of vector spaces
\bat{3}
\baf E\mul &\to \baf E\cen \car \f S\rel
&&:
(v_1, \dots, v_n)
&&\mto
\big(\sum_i \mu_i \, v_i\,, \;
(v_1 - \sum_i \mu_i \, v_i \,, \, \dots, \,
v_n - \sum_i \mu_i \, v_i)\big) \,.
\\
\baf E^*\mul &\to \baf E^*\cen \car \f S^*\rel
&&:(\alp_1, \dots, \alp_n)
&&\mto
\big(\sum_i \alp_i\,, \;
(\alp_1 - \mu_1 \, (\sum_i \alp_i) \,, \, \dots, \,
\alp_n - \mu_n \, (\sum_i \alp_i))\big) \,.
\end{alignat*}

These splittings turn out to be affine fibred splittings over
$\baf T$
orthogonal with respect to the rescaled metric
$G\mul \,.$\END
\ePr

The multi--metric
$g\mul$
splits into the product of a metric
$g\dia \seq g\cen$
of
$\f E\dia \seq \f E\cen$
and a metric
$g\rel$
of
$\f S\rel \,.$
We observe that
$g\cen = g\pat \,,$
in virtue of the equality
$\sum_i \mu_i = 1 \,.$
Therefore, the multi--metric
$G\mul$
splits into the product of the metric
$G\dia \seq G\cen = \tfr m\h g\pat$
of
$\f E\dia \seq \f E\cen$
and the metric
$G\rel = \tfr m\h g\rel$
of
$\f E\rel \,.$

The gravitational connection
$K\Nat\mul$
of the multi--spacetime
$\f E\mul$
splits into the product of a gravitational connection
$K\Nat\cen$
of
$\f E\cen$
and of a gravitational connection
$K\Nat\rel$
of
$\f S\rel \,.$
The connections
$K\Nat\cen$
and
$K\Nat\rel$
coincide with the connections induced by the affine structures of the
corresponding spaces (because affine isomorphisms between affine
spaces preserve the connections induced by the affine structures).
Moreover, the connections
$K\Nat\cen$
and
$K\Nat\rel$
preserve the metrics
$g\cen$
and
$g\rel \,.$
%--------------------------------------------------------------------%
\mysssec{Multi--electromagnetic field}
\label{Multi--electromagnetic field}
%--------------------------------------------------------------------%
\bsm
The splitting of the multi--spacetime yields a splitting of
the multi--electromagnetic field.
\esm

\bPr
The isomorphism
$\f E\cen \car \f S\rel \to \f E\mul$
yields a splitting of
$\C F\mul$
into the three components
\beq
\C F\mul =
\C F\mul\,\cen + \C F\mul\,\rel + \C F\mul\,\,\cen\,\rel \,,
\eeq
where
\bat{2}
\C F\mul\,\cen
&:
\f E\mul \to
(\B L^{1/2} \ten \B M^{1/2}) \ten \Lam^2T^*\f E\cen
&&\sub
(\B L^{1/2} \ten \B M^{1/2}) \ten \Lam^2T^*\f E\mul \,,
\\
\C F\mul\,\rel
&:
\f E\mul \to
(\B L^{1/2} \ten \B M^{1/2}) \ten \Lam^2T^*\f S\rel
&&\sub
(\B L^{1/2} \ten \B M^{1/2}) \ten \Lam^2T^*\f E\mul \,,
\\
\C F\mul\,\,\cen\,\rel
&:
\f E\mul \to
(\B L^{1/2} \ten \B M^{1/2}) \ten (T^*\f E\cen \wed T^*\f S\rel)
&&\sub
(\B L^{1/2} \ten \B M^{1/2}) \ten \Lam^2T^*\f E\mul \,,
\end{alignat*}
according to the following formula
\bal
\C F\mul\,\cen (e\mul; \, v\mul, w\mul)
&=
\sum_i \tfr {q_i}m F_i (e_i; \, v\cen\,_i, w\cen\,_i) \,,
\\
\C F\mul\,\rel(e\mul; \, v\mul, w\mul)
&=
\sum_i \tfr {q_i}m F_i (e_i; \, \ve v\rel\,_i, \ve w\rel\,_i) \,,
\\
\C F\mul\,\,\cen\,\rel(e\mul; \, v\mul, w\mul)
&=
\sum_i \tfr {q_i}m F_i (e_i; \, v\cen\,_i, \ve w\rel\,_i) +
\sum_i \tfr {q_i}m F_i (e_i; \, \ve v\rel\,_i, w\cen\,_i)
 \,,
\end{align*}
i.e.
\bal
\C F\mul\,\cen &(e\mul; \, v\mul, w\mul) =
\\
&=
- o(v\cen) \,
\sum_i \tfr{q_i}m \ve E_i[o] (e_i) \cdot \ve w\cen[o]
+ o(w\cen) \,
\sum_i \tfr{q_i}m \ve E_i[o] (e_i) \cdot \ve v\cen[o]
\\
&\quad+
\sum_i \tfr {q_i}m \ve
B_i(e_i) \, \cdot (\ve v\cen[o] \cro \ve w\cen[o]) \,,
\\
\C F\mul\,\rel&(e\mul; \, v\mul, w\mul) =
\sum_i \tfr {q_i}m \ve
B_i(e_i) \, \cdot (\ve v\rel\,_i \cro \ve w\rel\,_i)
\\
\C F\mul\,\,\cen\,\rel&(e\mul; \, v\mul, w\mul) =
\\
&=
- o(v\cen) \,
\sum_i \tfr{q_i}m \ve E_i[o] (e_i) \cdot \ve w\rel\,_i +
\sum_i \tfr{q_i}m
\ve B_i(e_i) \cdot (\ve w\rel\,_i \cro \ve v\cen [o]\big)
\\
&\quad+
o(w\cen) \,
\sum_i \tfr{q_i}m \ve E_i[o] (e_i) \cdot \ve v\rel\,_i -
\sum_i \tfr{q_i}m
\ve B_i(e_i) \cdot (\ve v\rel\,_i \cro \ve w\cen [o]) \,.
\end{align*}
for each
$e\mul \in \f E\mul$
and
$v\mul = v\cen + \ve v\rel \in \baf E\mul =
\baf E\cen + \f S\rel \,.$

On the other hand, the inclusion
$i\cen : \f E\cen \hto \f E\mul$
yields the scaled the 2--form
\beq
\C F\cen \byd i^*\cen\C F\mul = \C F\mul\,\cen \com i\cen :
\f E\cen \to (\B L^{1/2} \ten \B M^{1/2}) \ten \Lam^2T^*\f E\cen \,,
\eeq
given by
\beq
\C F\cen(e\cen; \, v\cen, w\cen) =
\tfr qm F\pat(e\cen; \, v\cen, w\cen)
\,.
\eeq

If
$q_i = k \, m_i$
and
$F\pat$
is spacelikely affine, then
$\C F\mul\,\,\cen \seq \C F\cen$
and
$\C F\mul\,\,\cen\,\rel = 0 \,.$\END
\ePr

\bPr
The potential
$\C A\mul$
for
$\C F\mul$
splits as
\beq
\C A\mul = \C A\cen + \C A\rel \,,
\sep{where}
\C A\cen : \f E\mul \to T^*\f E\cen \,,
\quad
\C A\rel : \f E\mul \to T^*\f E\rel \,,
\eeq
with
$\C A\cen(e\mul; \, v\mul) =
\sum_i \tfr {q_i}m A_i\big(e_i; \, v\cen\,_i\big)$
and
$\C A\rel(e\mul; \, v\mul) =
\sum_i \tfr {q_i}m A_i\big(e_i; \, v\rel\,_i\big) \,.$\END
\ePr

We stress that, in general, each of the three components of the
multi--electromagnetic field depends on the whole multi--spacetime and
not just on the corresponding components.
Hence, in general, the joined multi--connection
$K\mul$
does not split into the product of a joined multi--connection
$K\cen$
of
$\f E\cen$
and of a multi--connection
$K\rel$
of
$\f E\rel \,.$
As a consequence, in general, the equation of motion of the
multi--particle splits into a system of equations for the motion of
the center of mass and for the relative multi--motion, which are
coupled.

However, in the particular case when the pattern electromagnetic
field $F\pat$ is constant and the charges are proportional to the
masses (i.e., $q_i = k \, m_i$) the mixed term $\C F\cen\,\rel$
vanishes. In this case, the rescaled multi--electromagnetic field
$\C F\mul$ splits truly with respect to the two components of the
multi--spacetime
$\f E\cen$
and
$\f S\rel \,.$
Therefore, also the joined multi--connection splits with respect to
$\f E\cen$ and $\f S\rel \,.$ Hence, the equation of motion of the
multi--particle splits into a decoupled system of equations for the
motion of the center of mass and for the relative multi--motion.
%--------------------------------------------------------------------%
\subsection{Rigid body mechanics}
\label{Rigid body mechanics}
%--------------------------------------------------------------------%
\bsm
Finally, we achieve the scheme for a rigid body in the
framework of ``covariant classical mechanics'', by considering a
space-like rigid constraint on the multi--spacetime and assuming
as spacetime for the rigid body the constrained subbundle of the
multi--spacetime, which is called {\em rigid body spacetime\/}.
All objects related to this rigid--spacetime are called {\em
rigid--body objects\/} and labelled by the subscript 
$``\rig" \,.$
\esm
%--------------------------------------------------------------------%
\mysssec{Configuration space}
\label{Rigid body: Configuration space}
%--------------------------------------------------------------------%
To carry on our analysis, we need a `generalised' definition of
affine space.
Namely, we define a {\em generalised affine space\/} to be a triple
$(A,G,\cdot) \,,$
where
$A$
is a set,
$G$
a group and
$\cdot$
a transitive and free left action of
$G$ on $A \,.$
Note that, for every
$a \in A \,,$
the `left translation'
$L(a) : G \to A : g \mto ga$
is a bijection.

The generalised affine space $A$ is naturally parallelisable as
$TA = A \car \F g \,,$
where
$\F g$ is the Lie algebra of $G \,.$

\myskip

We consider a set
$\{l_{ij} \in \B L^2 \;\; | \;\; i,j = 1, \dots , n, \;\;
i \neq j, \;\; l_{ij} = l_{ji}, \;\; l_{ik} \leq l_{ij} + l_{jk}\}$
and define  the subsets
\bat{4}
i\rig
&: \f E\rig \; \byd
&&\{e\mul \in \f E\mul
&&\;\;|\;\; \| e_i - e_j\| = l_{ij},
\; 1 \leq i < j \leq n\}
&&\hto \f E\mul \,,
\\
i\rot
&: \f S\rot \byd
&&\{v\rel \in \f S\rel
&&\;\;|\;\; \| v_i - v_j \| = l_{ij},
\; 1 \leq i < j \leq n \}
&&\hto \f S\rel \,.
\end{alignat*}

We set
$\f E\rot \byd \f T \car \f S\rot \,.$

We stress that the rigid constraint does not affect the center of
mass.

The inclusion
$i\rig$
turns out to be equivariant with respect to the left action
of\linebreak
$O(\f S\pat, g\pat) \,,$
because the rigid constraint is invariant with respect to this
group.

Then, the spacelike orthogonal affine splitting
$\f E\mul = \f E\cen \ucar{\f T} \f E\rel =
\f E\cen \car \f S\rel$
restricts to a splitting
\beq
\f E\rig = \f E\cen \ucar{\f T} \f E\rot =
\f E\cen \car \f S\rot \,.
\eeq

Thus, we obtain a curved fibred  manifold
$t\rig : \f E\rig \to \f T$
consisting of the fibred product over
$\f T$
of the affine bundles
$t\cen : \f E\cen \to \f T$
and
$t\rot : \f E\rot \to \f T \,,$
or, equivalently, consisting of the Cartesian product
of the affine bundle
$t\cen : \f E\cen \to \f T$
with the spacelike submanifold
$\f S\rot \sub \f S\rel \,.$

\myskip

The 1st jet space of
$\f E\rig$
splits as
$J_1\f E\rig \seq
(\f E\cen \car \f U\cen) \car (\B T^* \ten T\f S\rot) \,.$

Each rigid observer
$o\pat : \f E\pat \to J_1\f E\pat \,,$
induces an observer
$o\rig : \f E\rig \to J_1\f E\rig \,.$
In particular, each inertial observer
$o\pat \in \f U\pat$
induces an observer
$o\rig \in \f U\cen \,,$
which is still called {\em inertial\/}.

\myskip

The inclusion
$i\rig$
yields the scaled spacelike Riemannian metric
\beq
g\rig \byd
i^*\rig \, g\mul : T\f E\rig \ucar{\f E\rig} T\f E\rig
\to \B L^2 \ten \Rn \,.
\eeq

\myskip

In order to further analyse the geometry of 
$\f E\rig \,,$ 
it suffices to study 
$\f S\rot \,.$
%--------------------------------------------------------------------%
\mysssec{Rotational space}
\label{Rotational space}
%--------------------------------------------------------------------%
The geometry of
$\f S\rot$
depends on the initial mutual positions of particles and is time
independent.
In particular, particles can either lie on a straight line, or lie on
a plane, or ``span'' the whole space. This can be formalised as
follows.

For each
$r\rot \in \f S\rot \,,$
let us consider the vector space
\beq
\langle r\rot\rangle \byd
\Span\big{\{}(r_i - r_j) \mid 1 \leq i <j \leq n\big{\}} \sub \f S\pat
\,.
\eeq

We can prove that the dimension of this space depends only on
$\f S\rot$
and not on the choice of
$r\rot \in \f S\rig \,.$
We call this invariant number
$c\rot$
the \emph{characteristic} of
$\f S\rot \,.$
We can have
$c\rot = 1, 2, 3 \,.$
We say that
$\f S\rot$
is \emph{strongly non degenerate} if
$c\rot = 3 \,,$
\emph{weakly non degenerate} if
$c\rot = 2 \,,$
\emph{degenerate} if
$c\rot = 1 \,.$

We observe that the natural actions of
$O(\f S\pat, g\pat)$
on
$\f S\rel$
restricts to
$\f S\rot \,.$

The inclusion
$i\rot$
turns out to be equivariant with respect to the left action of
$O(\f S\pat, g\pat) \,,$
because the rigid constraint is invariant with respect to this group.

The action of
$O(\f S\pat, g\pat)$
on
$\f S\rot$
is transitive.

For each
$v\rot \in \f S\rot \,,$
let us call
$H[r\rot] \sub O(\f S\pat, g\pat)$
the corresponding isotropy subgroup.

We can see that:

- in the {\em strongly non degenerate case\/} the isotropy subgroup
$H[r\rot ]$
is the trivial subgroup
$\{1\} \,;$

- in the {\em weakly non degenerate case\/} the isotropy subgroup
$H[r\rot ]$
is the discrete subgroup of reflections with respect to
$\langle v\rot \rangle\,;$

- in the {\em degenerate case\/} the isotropy subgroup
$H[r\rot ]$
is the 1 dimensional subgroup of rotations whose axis is
$\langle r\rot \rangle\,;$
we stress that this subgroup is not normal.

Hence, we can prove that:

-- $\f S\rot$
is {\em strongly non degenerate\/} if and only if the action of
$O(\f S\pat, g\pat)$
on
$\f S\rot$
is free;

-- $\f S\rot$
is {\em weakly non degenerate\/} if and only if the action of
$O(\f S\pat, g\pat)$
on
$\f S\rot$
is not free, but the action of
$SO(\f S\pat, g\pat)$
on
$\f S\rot$
is free;

-- $\f S\rot$
is {\em degenerate\/} if and only if the action of
$SO(\f S\pat, g\pat)$
on
$\f S\rot$
is not free.

Of course, if
$n = 2 \,,$
then
$\f S\rot$
is degenerate; if
$n = 3 \,,$
then
$\f S\rot$
can be degenerate or weakly non degenerate.

Furthermore, we can prove that:

-- if
$\f S\rot$
is {\em strongly non degenerate\/}, then
$\f S\rot$
is an affine space associated with the group
$O(\f S\pat, g\pat) \,;$

-- if
$\f S\rot$
is {\em weakly non degenerate\/}, then
$\f S\rot$
is an affine space associated with the group
$SO(\f S\pat, g\pat) \,;$

-- if
$\f S\rot$
is {\em degenerate\/}, then
$\f S\rot$
is a homogeneous manifold with two possible distinguished
diffeomorphisms (depending on a chosen orientation on the
straight line of the rigid body) with the unit sphere
$S^2(\B L^* \ten \f S\pat, g\pat) \,.$

So, the choice of a configuration
$v\rot \in \f S\rot$
and of a scaled orthonormal basis in
$\f S\pat \,,$
respectively,
yields the  following diffeomorphisms (via the action of
$O(\f S\pat, g\pat)$
on
$\f S\rot$)
\bat{3}
\f S\rot
&\seq
O(\f S\pat, g\pat)
&&\seq
O(3) \,,
&&\quad
\text{in the strongly non degenerate case;}
\\
\f S\rot
&\seq
SO(\f S\pat, g\pat)
&&\seq
SO(3) \,,
&&\quad
\text{in the weakly non degenerate case;}
\\
\f S\rot
&\seq
S^2\pat
&&\seq
S^2 \,,
&&\quad
\text{in the degenerate case,}
\end{alignat*}
where
$S^2\pat \sub \B L^* \ten \f S\pat$
is the unit sphere with respect to the metric
$g\pat \,.$

\myskip

From now on, for the sake of simplicity and for physical reasons
of continuity, in the non degenerate case, we shall refer only to
one of the two connected components of 
$\f S\rot \,.$ 
Accordingly, we shall just refer to the non degenerate case (without
specification of strongly or weakly non degenerate) as to the
degenerate cases.

\bPr\label{universal covering of rotational space} 
In the non degenerate case, by considering the isomorphism
$\f S\rot \seq SO(3) \,,$ 
and the well known two--fold universal covering 
$S^3 \seq SU(2) \to SO(3) \,,$ 
we obtain the universal covering 
$S^3 \to \f S\rot \,,$ 
which is a principal bundle associated with the
group 
$\B Z_2$ \cite[vol.1]{KobNom69}.\END 
\ePr

This is in agreement with the fact that the homotopy group of
$\f E\rot$
is
\cite[vol.2]{DubNovFom79}
\beq
\pi_1(\f S\rot) =
\pi_1\big((SO(\f S\pat, g\pat)\big) = \B Z_2 \,.
\eeq
%--------------------------------------------------------------------%
\mysssec{Tangent space of rotational space}
\label{Tangent space of rotational space}
%--------------------------------------------------------------------%
\mypar{Non degenerate case.}
\label{Non degenerate case}
%--------------------------------------------------------------------%
The generalised affine structure of
$\f S\rot \,,$
with respect to the group
$O(\f S\pat, g\pat) \,,$
yields the natural parallelisation
\beq
T\f S\rot = \f S\rot \car \F {so}(\f S\pat, g\pat) \,.
\eeq

We can regard this isomorphism in another interesting way, which
expresses in a geometric language the classical formula of velocity
of a rigid body.

\myskip

For this purpose, let us consider the three dimensional scaled vector
space
\beq
\f V\anl \byd \B L^* \ten \f S\pat \,.
\eeq

Then, the metric
$g\pat$
and the chosen orientation of
$\f S\pat$
determine the linear isomorphisms
\beq
g\Fla : \F {so}(\f S\pat, g\pat) \to \B L^{2}
\ten \Lam^2 \f S\pat^*
\ssep{and}
* : \B L^{2} \ten \Lam^2 \f S\pat^* \to \f V\anl \,,
\eeq
hence the linear isomorphism
\beq
\F {so}(\f S\pat, g\pat) \seq \f V\anl \,.
\eeq

Therefore, we can read the above parallelization also as
\bEq
\tau\anl : T\f S\rot \seq \f S\rot \car \f V\anl \,.
\eEq

The inverse of the above isomorphism
\bEq
\tau^{-1}\anl : \f S\rot \car \f V\anl \to T\f S\rot
\sub \f S\rot \car \f S\rel
\eEq
is expressed by the formula
\beq
(r_1, \dots, r_n \,; \ome) \mto
(r_1, \dots, r_n \,; \;\; \ome \cro r_1, \dots, \ome \cro r_1)
\,,
\eeq
where
$\cro$
is the cross product of
$\f S\pat$
defined by
$u \cro v \byd g\Sha\pat (i(u \wed v) \, \eta\pat) \,,$
where
$\eta\pat$
is the metric volume form of
$\f S\pat \,.$
The above formula is just a geometric formulation of the well known
formula expressing the relative velocity of the particles of a rigid
body through the angular velocity.

Thus, for each
$(r_1, \dots, r_n \,,\; v_1, \dots, v_n) \in T\f S\rot
\sub
\f S\rot \car \f S\rel \,,$
there is a unique
$\ome \in \f V\anl$
such that
$v_i = \ome \cro r_i \,,$
for
$1 \leq i \leq n \,.$

The cross product
$\cro$
of
$\f S\pat$
is equivariant with respect to the left action of
$SO(\f S\pat, g\pat) \,.$
Hence, the isomorphism
$\tau\anl$
turns out to be equivariant with respect to this group.

The {\em angular velocity\/} of a rigid motion
$s : \f T \to \f E\rig$
is defined to be the map
\beq
\ome \byd \tau\anl \com T\pi\rot \com ds :
\f T \to \B T^* \ten \f V\anl \,,
\eeq
where
$\pi\rot : \f E\rig \to \f S\rot$
is the natural projection map according to section
\ref{Rigid body: Configuration space}.

We stress that the above geometric constructions use implicitly
the pattern affine structure. Hence, the angular velocity is
independent of the choice of inertial observers. But, the observed
angular velocity would depend on the choice of non inertial
observers.
%--------------------------------------------------------------------%
\mypar{Degenerate case.}
\label{Degenerate case}
%--------------------------------------------------------------------%
According to a well--known result on homogeneous spaces, the tangent
space of
$\f S\rot$
turns out to be the quotient vector bundle
\beq
T\f S\rot = \f S\rot \car \F{so}(\f S\pat, g\pat) / h[\f S\rot] \,,
\eeq
where
$h[\f S\rot] \sub \f S\rot \car \F{so}(\f S\pat, g\pat)$
is the vector subbundle
over
$\f S\rot$
consisting of the isotropy Lie algebras of
$\f S\rot \,.$

Now, let us consider again the scaled vector space
$\f V\anl \byd \B L^* \ten \f S\pat$
and define the quotient vector bundle over
$\f S\rot$
\beq
(\f S\rot \car \f V\anl)/\sim \,,
\eeq
induced, for each
$r\rot \in \f S\rot \,,$
by the vector subspace
$\langle r\rot\rangle \sub \f V\anl$
generated by
$r\rot \,.$

Then, by proceeding as in the non degenerate case and taking the
quotient with respect to the isotropy subbundle, we obtain the
linear fibred isomorphism 
\beq 
[\tau\anl] : 
T\f S\rot \seq (\f S\rot \car \f V\anl)/\sim \,. 
\eeq

The inverse of the above isomorphism 
\beq 
[\tau\anl]^{-1} : (\f S\rot \car \f V\anl)/ \sim \; \to \; T\f S\rot
\sub \f S\rot \car \f S\mul
\eeq 
is expressed by the formula 
\beq 
(r_1, \dots, r_n
\,; [\ome]) \mto (r_1, \dots, r_n \,; \;\; 
\ome \cro r_1, \dots, \ome \cro r_1) \,, 
\eeq 
where the cross products 
$\ome \cro r_i$
turns out to be independent on the choice of representative for
the class 
$[\ome] \,.$

Thus, for each
$(r_1, \dots, r_n \,,\; v_1, \dots, v_n) \in T\f S\rot
\sub
\f S\rot \car \f S\rel \,,$
there is a unique
$[\ome] \in (\f S\rot \car \f V\anl)/ \sim_{|(r_1, \dots, r_n)}$
such that
$v_i = \ome \cro r_i \,,$
for
$1 \leq i \leq n \,.$

Clearly, each choice of the orientation of the rigid body yields a
distinguished fibred isomorphism
\beq
T\f S\rot \seq TS^2(\B L^* \ten \f S\pat, g\pat) \,.
\eeq
%--------------------------------------------------------------------%
\mysssec{Induced metrics}
\label{Induced metrics}
%--------------------------------------------------------------------%
\bsm
The multi--metric of
$\f S\mul$
induces a metric on
$\f S\rot \,,$
which can be regarded also in another useful way through the
isomorphism
$\tau\anl \,.$

Even more, the standard pattern metric of 
$\f V\anl$ 
induces a further metric on 
$\f S\rot \,,$ 
which will be interpreted as the inertia tensor.
\esm

The inclusion
$i\rot$
yields the scaled Riemannian metric
\beq
g\rot \byd
i^*\rot \, g\rel :
T\f S\rot \ucar{\f S\rot} T\f S\rot
\to \B L^2 \ten \Rn \,.
\eeq

\myskip

We can regard this metric in another interesting way, which follows
from the parallelisation through
$\f V\anl \,.$

For this purpose, the patter metric
$g$
can be regarded as a Euclidean metric
of
$\f S\pat$
\beq
g : \f V\anl \car \f V\anl \to \Rn \,.
\eeq

We can make the natural identifications
$O(\f V\anl, g) \seq O(\f S\pat, g\pat) \,.$

\myskip

Therefore, the isomorphism
$\tau\anl$
allows us to read
$g\rot$
as the scaled fibred Riemmannian metric
\bal
\sig
&\byd \tau^{-1*}\anl \, g\rot :
\f S\rot \car (\f V\anl \car \f V\anl)
\to \B L^2 \ten \Rn
\\
\sig
&\byd \tau^{-1*}\anl \, g\rot :
\big(\f S\rot \car \f V\anl)/\sim\big) \ucar{\f S\rot}
\big(\f S\rot \car \f V\anl)/\sim\big)
\to \B L^2 \ten \Rn \,,
\end{align*}
respectively, in the non degenerate and in the degenerate cases.
Its expression is
\bAt{2}\label{expression of the inertia tensor}
\sig
&(r_1, \dots, r_n\,; \; \ome, \, \ome')
&&=
\sum_i \mu_i \, \big(
g\pat(r_i, r_i) \, g\pat(\ome,\ome') -
g\pat(r_i, \ome) \, g\pat(r_i, \ome')
\big)
\\ \nonumber
\sig
&(r_1, \dots, r_n\,; \; [\ome], \, [\ome'])
&&=
\sum_i \mu_i \, \big(
g\pat(r_i, r_i) \, g\pat(\ome,\ome') -
g\pat(r_i, \ome) \, g\pat(r_i, \ome')
\big) \,,
\end{alignat}
respectively, in the non degenerate and in the degenerate cases.

In the degenerate case, the above expression can be also written as
\beq
\sig (r_1, \dots, r_n\,; \; [\ome], \, [\ome']) =
g\pat(\ome,\ome') \, \sum_i \mu_i \, g\pat(r_i, r_i) \,,
\eeq
where
$\ome$
and
$\ome'$
are the representatives of
$[\ome]$
and
$[\ome']$
orthogonal to the
$r_i$'s \,.

\myskip

Then, we obtain a further metric.
In fact, the metric
$g$
of
$\f V\anl$
can be regarded as a fibred metric over
$\f S\rot \,,$
which will be denoted by the same symbol,
\bal
g
&:
\f S\rot \car (\f V\anl \car \f V\anl) \to \Rn
\\
g
&:
\big(\f S\rot \car \f V\anl)/\sim\big) \ucar{\f S\rot}
\big(\f S\rot \car \f V\anl)/\sim\big) \to \Rn \,,
\end{align*}
respectively, in the non degenerate and in the degenerate cases,
according to the equalities
\bal
g (r_1, \dots, r_n\,; \; \ome, \, \ome')
&=
g\pat (\ome, \ome')
\\
g (r_1, \dots, r_n\,; \; [\ome], \, [\ome'])
&=
g\pat (\ome\per, \ome'\per) \,,
\end{align*}
where 
$\ome\per$ 
and 
$\ome'\per$ 
are the components of
$\ome$ 
and 
$\ome'$ 
orthogonal to 
$r_i \,.$

Then, we obtain the further unscaled Riemannian metric of
$\f S\rot$
\beq
\sig\rot \byd \tau^*\anl \, g :
T\f S\rot \ucar{\f S\rot} T\f S\rot \to \Rn \,.
\eeq

\myskip

All metrics of $\f S\rot$ considered above are invariant with
respect to the left action of $O(\f S\pat, g\pat) \,.$

\bPr\label{metric isomorphisms in the non degenerate case} In the
non degenerate case, the choice of a configuration 
$r\rot \in \f S\rot$ 
and of a scaled orthonormal basis in 
$\f V\anl \,,$
respectively, yields the  following diffeomorphisms (via the
action of 
$SO(\f V\anl, g\anl)$ 
on 
$\f S\rot$)
\beq
\f S\rot \seq SO(\f V\anl, g\anl) \seq SO(3) \,,
\eeq
which turn out to be isometries with respect to the Riemannian metrics
$\sig\rot \,,$
$- \tfr12 k\anl$ and $-\tfr12 k_3 \,,$ of $\f
S\rot \,,$ $\f V\anl$ and $SO(3) \,,$ 
where 
$k\anl$ 
and 
$k_3$ 
are the Killing metrics.
\ePr

\bpf
The above diffeomorphisms yield the linear fibred isomorphisms
\beq
T\f S\rot \seq \F{so}(\f V\anl, g\anl) \seq \F{so}(3) \,.
\eeq

On the other hand, the natural isomorphism
$\F{so}(\f V\anl, g\anl) \to \f V\anl$
is metric.
Hence, in virtue of the definition of
$\sig\rot \,,$
the isomorphism
$T\f S\rot \seq \F{so}(\f V\anl, g\anl)$
turns out to be metric.

Moreover, the metric 
$g\anl$ 
of 
$\f V\anl$ 
turns out to coincide with the metric
$- \tfr12 k\anl$
of
$\F{so}(\f V\anl, g\anl) \,.$ 
In fact, we have
$g\anl(\ome, \ome') =
- \tfr12 \, \tr \big((\ome \cro) \com (\ome' \cro)\big) \,.$
By a standard argument, the isomorphism \linebreak
$\F{so}(\f V\anl, g\anl) \seq \F{so}(3)$
turns out to be metric.\QED
\epf

In a similar way, we can prove the following result.

\bPr\label{metric isomorphisms in the degenerate case}
In the degenerate case, the choice of a configuration
$r\rot \in \f S\rot$
and of a scaled orthonormal basis in
$\f V\anl \,,$
respectively,
yields the  following diffeomorphisms (via the action of
$SO(\f V\anl, g)$
on
$\f S\rot$)
\beq
\f S\rot \seq S^2\anl \seq S^2 \,,
\eeq
which turn out to be isometries with respect to the metrics
$\sig\rot$
of
$\f S\rot \,,$
the metric
$g$
of
$S^2\anl$
(induced by the inclusion
$S^2\anl \sub \f V\anl$)
and the metric
$g_2$
of
$S^2$
(induced by the inclusion
$S^2 \sub \Rn^3$).\END
\ePr
%--------------------------------------------------------------------%
\mypar{Inertia tensor.}
\label{Inertia tensor}
%--------------------------------------------------------------------%
The fibred metric
$g$
of
$\f S\rot$
allows us to regard the fibred metric
$\sig$
of
$\f S\rot$
as a scaled symmetric fibred automorphism
\beq
\hat\sig :
\f S\rot \to \B L^2 \ten (\f V^*\anl \car \f V\anl) \,.
\eeq

The scaled metric
$m \, \sig \,,$
or the scaled automorphism
$m \, \ha\sig \,,$
are called the {\em inertia tensor\/}.
The scaled eigenvalues of the inertia tensor are called {\em principal
inertia momenta\/} and are denoted by
$I_i \in \map(\f S\rot, \, \B L^2 \ten \B M \ten \Rn) \,.$
Indeed, the principal inertia momenta turn out to be constant with
respect to
$\f S\rot \,.$

In the non degenerate case, we have three principal inertia momenta.
Then, three cases can occur:
\bal
I \byd I_1 = I_2 = I_3 \,,
&\qquad
\text{\em spherical case} \,,
\\
I \byd I_1 = I_2 \neq I_3 \,,
&\qquad
\text{\em symmetric case} \,,
\\
I_1 \neq I_2 \neq I_3 \neq I_1 \,,
&\qquad
\text{\em asymmetric case}.
\end{align*}

In the degenerate case, we have two coinciding principal inertia
momenta
\beq
I \byd I_1 = I_2 = \sum_i m_i \, g\pat(r_i, r_i) \,.
\eeq

In the spherical non degenerate case and in the degenerate case, we
have
\bEq\label{proportionality of metrics}
g\rot = \fr Im \, \sig \,.
\eEq

\myskip

Thus, we have studied the diagonalisation of $\sig$ with
respect to 
$g \,.$ 
In an analogous way, we can diagonalise
$g\rot$ 
with respect to 
$\sig\rot \,.$ 
Indeed, in this way we obtain the same eigenvalues and the same
classification, because the two diagonalisations are related by the
isomorphism 
$\tau\anl \,.$

\myskip

The principal inertia momenta are related to the scalar curvature of
the rotational space in the following way.

\bPr
The scalar curvature of
$\f S\rot \,,$
with respect to the metric
$G\rot \,,$
is
\cite{Tej01}
\bat{2}
\rho\rot
&= \fr{3 \, \h}{2 \, I} \,,
&&\quad
\text{sph. non deg. case, with }
I \byd I_1 = I_2 = I_3 \,,
\\
\rho\rot
&= \fr{2 \, \h}{I_1} - \fr{\h \, I}{2 \, I^2_1} \,,
&&\quad
\text{sym. non deg. case, with }
I \byd I_2 = I_3 \,,
\\
\rho\rot
&=
\fr\h{I_1} + \fr\h{I_2} + \fr\h{I_3}
- \fr{\h \, (I_1^2 + I_2^2 + I_3^2)}{2 \, I_1 \, I_2 \, I_3} \,,
&&\quad
\text{asym. non deg. case,}
\\
\rho\rot
&= \fr{2 \, \h}{I} \,,
&&\quad
\text{deg. case, with }
I \byd I_1 = I_2 \,.
\end{alignat*}

Moreover, since the splitting
$\f E\rig = \f E\cen \car \f S\rot$
is orthogonal, the vanishing of the scalar curvature
$\rho$ 
yields 
$\rho\rig = \rho\rot \,.$\END
\ePr
%--------------------------------------------------------------------%
\mysssec{Induced connection}
\label{Induced connection}
%--------------------------------------------------------------------%
\bsm
The multi--connection of the multi--spacetime induces naturally a
connection on the rigid configuration space, which splits naturally
into the center of mass and relative components.
\esm

We can easily state the following generalisation of a well known
theorem due to Gauss
\cite{KobNom69}.

\bLm
Let us consider a fibred manifold
$p : \f F \to \f B$
equipped with a vertical Riemannian metric
$g_\f F$
and a linear connection
$K_\f F$
of
$\f F \,,$
which restricts to the fibres of
$\f F \to \f B$
and preserves the metric
$g_\f F \,.$

Moreover, let us consider a fibred submanifold
$\f G \sub \f F$
over
$\f B$
and the orthogonal projection
$\pi_\f G : T\f F_{|\f G} \to T\f G$
induced by
$g_\f F \,.$

Then, there exists a unique linear connection 
$K_\f G$ 
of 
$\f G \,,$ 
which restricts to the fibres of 
$\f G \to \f B$ 
and such that, for every pair of vector fields 
$X, \, Y$ of $\f G \,,$ 
we have 
$\pi_\f G (\nab[K_\f F]_X Y) = \nab[K_\f G]_X Y \,.$
Moreover, this connection 
$K_\f G$ 
preserves 
$g_\f G \,.$\END
\eLm

According to the above Lemma, the connection
$K\Nat\mul$
of
$\f E\mul$
yields a linear connection
$K\Nat\rig$
of
$\f E\rig \,,$
which preserves the time fibring and the metric
$g\rig \,.$

Moreover, according to a standard result due to Gauss, the
connection 
$K\rel$ 
of 
$\f S\rel$ 
induces a connection
$\vkap\Nat\rot$ 
on 
$\f S\rot \,,$ 
which coincides with the Riemannian connection induced by 
$g\rot \,.$

\bPr
By considering the splitting
$\f E\rig = \f E\cen \car \f S\rot \,,$
the connection
$K\Nat\rig$
splits into the product of the connections
$K\Nat\cen$
and
$\vkap\Nat\rot \,.$
\ePr

\bpf
We have the splitting
$K\Nat\mul = K\Nat\cen \car K\Nat\rel \,.$
Moreover, the splitting
$\f E\mul = \f E\cen \car \f S\rel$
is orthogonal with respect to the metric
$g\mul \,,$
hence the projection
$\pi_{\f E\rig}$
splits into the projections
$\f E\mul \to \f E\cen$
and
$\f E\mul \to \f S\rel \,.$

Hence,
$K\Nat\rig$
splits into the product of the connections
$K\Nat\cen$
and
$\vkap\rot \,.$\QED
\epf
%--------------------------------------------------------------------%
\mysssec{Induced electromagnetic field}
\label{Induced electromagnetic field}
%--------------------------------------------------------------------%
\bsm
We analyse the pullback of the multi electromagnetic field on the
rigid spacetime.
This is a 2--form on a 1+6 dimensional manifold in the non degenerate
case and on a 1+5 dimensional manifold in the degenerate case.
We can express this 2--form in terms of the pattern electric and
magnetic fields.

We can decompose this form into three components: the center of
mass component, the rotational component and the mixed component.
In the particular case when the mixed component vanishes and the
other two components depend only on the center of mass and
rotational variables, these two components coincide with the
pullback of the multi electromagnetic field with respect to the
center of mass and rotational projections.

Indeed, we can prove that the pullback of the multi--electromagnetic
field on the rigid spacetime provides the suitable electromagnetic
object for the correct expression of the classical law of motion (in
the context of our formulation of classical mechanics of a rigid body
interpreted as a classical particle moving in a higher dimensional
spacetime).

Therefore, we shall assume this pullback also as the correct
object for our formulation of quantum mechanics of a rigid body.
\esm
%--------------------------------------------------------------------%
\mypar{Non degenerate case.}
%--------------------------------------------------------------------%
Let us start by studying the non degenerate case.

\myskip

\bPr
The inclusion
$i\rig : \f E\rig = \f E\cen \car \f S\rot \hto \f E\mul$
yields the scaled 2--form
\beq
\C F\rig \byd i^*\rig \C F\mul \,,
\eeq
which splits into the three components
\beq
\C F\rig =
\C F\rig\,\cen + \C F\rig\,\rot + \C F\rig\,\,\cen\,\rot \,,
\eeq
where
\bal
\C F\rig\,\cen &: \f E\rig \to
(\B L^{1/2} \ten \B M^{1/2}) \ten \Lam^2T^*\f E\cen
\\
\C F\rig\,\rot &: \f E\rig \to
(\B L^{1/2} \ten \B M^{1/2}) \ten \Lam^2T^*\f S\rot
\\
\C F\rig\,\,\cen\,\rot &: \f E\rig \to
(\B L^{1/2} \ten \B M^{1/2}) \ten (T^*\f E\cen \wed T^*\f S\rot) \,,
\end{align*}
according to the following formula
\bal
\C F\rig\,\cen (e\rig; \; v\rig, w\rig)
&=
\sum_i \tfr {q_i}m F_i\big(e_i; \; v\cen\,_i, w\cen\,_i\big)
\\
\C F\rig\,\rot(e\rig; \, v\rig, w\rig)
&=
\sum_i \tfr {q_i}m F_i
\big(e_i; \;
\ome \cro r_i, \, \psi \cro r_i\big)
\\
\C F\rig\,\,\cen\,\rot(e\rig; \, v\rig, w\rig)
&=
\sum_i \tfr {q_i}m
F_i\big(e_i; \; v\cen\,_i, \, \psi \cro r_i\big) +
\sum_i \tfr {q_i}m
F_i\big(e_i; \; \ome \cro r_i, \, w\cen\,_i\big) \,,
\end{align*}
for each
\bal
(e\rig, v\rig)
&=
(e\cen, \, r_1, \dots, r_n \,; \;
v\cen +
\ome \cro r_1 + \dots + \ome \cro r_n) \in T\f E\rig \,,
\\
(e\rig, w\rig)
&=
(e\cen, \, r_1, \dots, r_n \,; \;
w\cen +
\psi \cro r_1 + \dots + \psi \cro r_n) \in T\f E\rig \,,
\end{align*}
i.e.
\bal
\C F\rig\,\cen (e\rig; \; v\rig, w\rig)
&=
\sum_i \tfr {q_i}m \ve E_i[o](e_i)
\cdot (\ve v\cen[o] - \ve w\cen[o])
\\
&+
\sum_i \tfr {q_i}m \ve B_i(e_i) \cdot
\big(\ve v\cen[o] \cro \ve w\cen[o]\big)
\\
\C F\rig\,\rot (e\rig; \; v\rig, w\rig)
&=
\sum_i \tfr {q_i}m \big(\ve B_i(e_i)
\cdot r_i\big) \,
\big((\ome \cro \psi) \cdot r_i\big)
\\
\C F\rig\,\,\cen\,\rot (e\rig; \; v\rig, w\rig)
&=
\sum_i \tfr {q_i}m \ve E_i[o] (e_i) \cdot
\big((\ome - \psi) \cro r_i\big)
\\
&+ \sum_i \tfr {q_i}m \big(\ve B_i (e_i) \cdot
\psi\big) \,
\big(\ve v\cen[o] \cdot r_i\big)
\\
&- \sum_i \tfr {q_i}m \big(\ve B_i (e_i) \cdot
r\rot\,_i\big) \,
\big(\ve v\cen[o] \cdot \psi\big)
\\
&- \sum_i \tfr {q_i}m \big(\ve B_i (e_i) \cdot
\ome\big) \,
\big(\ve w\cen[o] \cdot r\rot\,_i\big)
\\
&+ \sum_i \tfr {q_i}m \big(\ve B_i (e_i\big) \cdot
r\rot\,_i\big) \,
\big(\ve w\cen[o] \cdot \ome\big) \,.
\end{align*}

If
$q_i = k \, m_i$
and
$F\pat$
is spacelikely affine, then
$\C F\rig\,\cen \seq \C F\cen$
and
$\C F\rig\;\cen\,\rot = 0 \,.$\END
\ePr

\bPr
The potential
$\C A\rig \byd i^*\rig \C A\mul$
for
$\C F\rig$
splits as
\beq
\C A\rig =
\C A\rig\,\cen + \C A\rig\,\rot \,,
\sep{where}
\C A\rig\,\cen : \f E\rig \to T^*\f E\cen \,,
\quad
\C A\rig\,\rot : \f E\rig \to T^*\f S^* \,,
\eeq
with
$\C A\cen(e\rig; \, v\rig) =
\sum_i \tfr {q_i}m A_i (e\rig\,_i; \, v\cen\,_i)$
and
$\C A\rel(e\rig; \, v\rig) =
\sum_i \tfr {q_i}m A_i (e\rig\,_i; \, v\rot\,_i) \,.$\END
\ePr
%--------------------------------------------------------------------%
\mypar{Degenerate case.}
%--------------------------------------------------------------------%
The degenerate case can be studied in a similar way to the non
degenerate one.

Here, we just provide, as an example, an explicit description of a
{\em dipole\/}, consisting of 2 particles with opposite charges in
a constant electromagnetic field. In this case, we have 
\bal 
\C F\rig\,\cen (e_1, e_2; \, v\rig, w\rig) 
&= \sum_i \tfr{q_i}m
F(v\cen, w\cen) = 0
\\[2mm]
\C F\rig\,\rot (e_1, e_2; \, v\rig, w\rig)
&=
\tfr {q_1}m F (v\rot\,_1, w\rot\,_1) -
\tfr {q_1}m (\tfr{m_1}{m_2})^2 F (v\rot\,_1, w\rot\,_1)
\\
&=
q_1 \tfr{m_2 - m_1}{m^2_2} F (v\rot\,_1, w\rot\,_1)
\\
&=
2 \, q_1 \tfr{m_2 - m_1}{m^2_2} \ve B \cdot
(v\rot\,_1 \cro w\rot\,_1)
\\[2mm]
\C F\rig\;\cen\,\rot (e_1, e_2; v\rig, w\rig)
&=
\tfr{q_1}m F (v\cen\,_1, w\rot\,_1) -
\tfr{q_1}m F (v\cen\,_2, w\rot\,_2)
\\
&+
\tfr{q_1}m F (v\rot\,_1, w\cen\,_1) -
\tfr{q_1}m F (v\rot\,_2, w\cen\,_2)
\\[3mm]
&=
\tfr{q_1}m F (v\cen\,_1, w\rot\,_1) +
\tfr{m_1}{m_2} \tfr{q_1}m F (v\cen\,_1, w\rot\,_1)
\\
&+
\tfr{q_1}m F (v\rot\,_1, w\cen\,_1\big) +
\tfr{m_1}{m_2} \tfr{q_1}m F (v\rot\,_1, w\cen\,_1)
\\[3mm]
&=
2 \, \tfr{q_1}{m_2} \ve E\pat[o] \cdot
(w^0\cen \, v\rot\,_1) - v^0\cen \, w\rot\,_1))
\\
&+
2 \, \ve B\pat \cdot
g (v\rot\,_1 \cro \ve w\cen[o] + w\rot\,_1 \cro \ve v\cen[o]) \,.
\end{align*}
%--------------------------------------------------------------------%
\mysssec{Spacetime structures}
\label{Spacetime structures}
%--------------------------------------------------------------------%
\bsm
The previous results suggest a model for the classical
mechanics of a rigid body completely analogous to our one--body
scheme.
\esm
 
We assume the fibred manifold 
$t\rig : \f E\rig \to \f T$ 
as {\em rigid--body spacetime\/}. 
Moreover, we assume the metric
$g\rig \byd i^*\rig \, g\mul$ 
as the {\em spacelike metric\/}, the metric 
$G\rig = \tfr m\h g\rig$ 
as the {\em rescaled spacelike metric\/}, the connection 
$K\Nat\rig = K\Nat\cen \car \vkap\Nat\rig$ 
as the {\em gravitational connection\/}, the 2--form 
$\C F\rig \byd i^*\rig \, \C F\mul$ 
as the {\em rescaled electromagnetic field\/}, and the 2--form 
$\tfr m\h \, \C F\rig$
as the {\em unscaled electromagnetic field\/}.

The joined cosymplectic 2--form 
$\Ome\rig$ 
induced by the above gravitational connection, the unscaled
electromagnetic field and the rescaled metric coincides with the
pullback 
$\Ome\rig = i^* \Ome\mul \,.$

Hence,
$\Ome\rig$
turns out to be a globally exact cosymplectic 2--form.

\myskip

The velocity space of
$\f E\rig$
splits as
\beq
J_1\f E\rig \seq
(\f E\cen \car \f U\cen) \car (\B T^* \ten T\f S\rot) \,.
\eeq

\myskip

An inertial observer
$o$
yields the further splitting
$\f E\cen = \f T \car \f P[o]\cen \,.$

Given an inertial observer
$o \,,$
we shall refer to a spacetime chart
$(x^0, x^i, x^\alp)$
adapted to the observer and to the center of mass splitting.
Here, indices
$i,j$
will label coordinates of
$\f P[o]\cen$
and
$\alp, \bet$
will label coordinates of
$\f S\rot$
(e.g., Euler angles).
%--------------------------------------------------------------------%
\mysssec{Dynamical functions}
\label{Dynamical functions}
%--------------------------------------------------------------------%
\bsm
Here, we discuss the momentum and Hamiltonian functions and their
splitting into the translational and rotational components.
\esm

Let us choose a horizontal potential
$A\Upa\rig$
for
$\Ome\rig$
and an inertial observer
$o \,.$

They yield the rigid momentum and Hamiltonian
\beq
\C P\rig \byd \nu[o] \con A\Upa\rig : J_1\f E\rig \to T^*\f E\rig \,,
\qquad
\C H\rig \byd - o \con A\Upa\rig : J_1\f E\rig \to T^*\f E\rig \,,
\eeq
which split as
\beq
\C P\rig = \C P\cen + \C P\rot \,,
\qquad
\C H\rig = \C H\cen + \C H\rot \,,
\eeq
where
\bat{3}
\C P\cen
& : J_1\f E\rig \to T^*\f E\cen \,,
\qquad
&&\C P\rot
&& : J_1\f E\rig \to T^*\f E\rot \,,
\\
\C H\cen
&: J_1\f E\rig \to T^*\f E\cen \,,
\qquad
&&\C H\cen
&&: J_1\f E\rig \to T^*\f E\rot \,.
\end{alignat*}

We have the coordinate expressions
\bat{3}
\C P\cen\,_j
&=
G\cen\,^0_{ij} \, x^j_0 + A\cen\,_i \,,
\qquad
&&\C P\rot\,_\alp
&&= G\rot\,^0_{\alp\bet} \, x^\bet_0 + A\rot\,_\alp \,,
\\
\C H\cen\,_0
&=
\tfr12 G\cen\,^0_{ij} \, x^i_0 \, x^j_0 - A\cen\,_0 \,,
\qquad
&&\C H\rot\,_0
&&=
\tfr12 G\rot\,^0_{\alp \bet} \, x^\alp_0 \, x^\bet_0 - A\rot\,_0
\,.
\end{alignat*}

Clearly, 
$\C P\cen$ 
and 
$\C P\rot$ 
can be identified with the angular momentum of the center of mass and
the angular momentum with respect to the center of mass, respectively.

In the general case they are coupled and not conserved.

In the particular case when
$F = 0 \,,$
they are conserved and we obtain the decoupled expressions
\bat{3}
\C P\cen
& : J_1\f E\cen \to T^*\f E\cen \,,
\qquad
&&\C P\rot
&& : J_1\f E\rot \to T^*\f E\rot \,,
\\
\C H\cen
&: J_1\f E\cen \to T^*\f E\cen \,,
\qquad
&&\C H\cen
&&: J_1\f E\rot \to T^*\f E\rot \,.
\end{alignat*}
%--------------------------------------------------------------------%
\newpage
\section{Rigid body quantum mechanics}
\label{Rigid body quantum mechanics}
%--------------------------------------------------------------------%
\bsm
In the previous chapter we have described the classical framework of
a rigid body in analogy with the framework of a constrained one--body.
Then, we approach the quantisation of the rigid body according
to the scheme of ``covariant quantum mechanics", by analogy with the
case of a one-body.

We define quantum structures, analyse their existence and classify
them.
Then, we evaluate the quantum operators and compute the spectra
of the energy operator in some cases.
\esm
%--------------------------------------------------------------------%
\subsection{Quantum structures}
\label{Quantum structures}
%--------------------------------------------------------------------%
First, we analyse the existence and classification of quantum
structures according to Proposition \ref{existence and
classification}.

The existence condition of the quantum structure is fulfilled due to
the exactness of
$\Ome\rig \,:$
\beq
[\Ome] = 0 \in
i^2 \big(H^2(\f E, \B Z)\big) \sub
H^2(\f E, \Rn) \seq
H^2(J_1\f E, \Rn) \,.
\eeq

So, we have just to compute all possible inequivalent quantum
structures.
%--------------------------------------------------------------------%
\mypar{Non degenerate case.} \label{Non degenerate caseII}
%--------------------------------------------------------------------%
Let us start with the non degenerate case.

\bPr
We have just two equivalence classes of complex line bundles over
$\f E\rig \,.$
Clearly, one of these classes is the trivial one.
Indeed, both of them admit quantum connections.
\ePr

\bpf
The 2nd cohomology groups of 
$\f E\rig$ 
are
\cite{BotTu82}:
\bat{3}
H^2(\f E\rig, \B Z) 
&\seq H^2(\f
S\rig, \B Z) 
&&\seq H^2\big(SO(\f S\pat, g\pat), \B Z\big) 
&&\seq
\B Z_2 \,,
\\
H^2(\f E\rig, \Rn)
&\seq
H^2(\f S\rig, \Rn)
&&\seq
H^2\big(SO(\f S\pat, g\pat), \Rn\big)
&&\seq
\{0\} \,.
\end{alignat*}

Then, according to
Proposition \ref{existence and classification},
the equivalence classes of complex line bundles are in bijection with
$H^2(\f E\rig, \B Z) = \B Z_2$
and the equivalence classes of quantum bundles are in bijection with
$(i^2)^{-1}([\Ome]) = (i^2)^{-1}(0) = \B Z_2 \,.$\QED
\epf

We can produce two concrete representatives for the above
equivalence classes of vector bundles in the following way.

\bLm\label{trivial and non trivial rotational quantum bundles}
The two inequivalent representations of
$\B Z_2$
on
$\B C$
yield the trivial Hermitian line bundle
$\f Q^+\rot$
and the non trivial Hermitian line bundle
$\f Q^-\rot \,,$
equipped with flat Hermitian connections
$\chi^+\rot$
and
$\chi^-\rot \,,$
respectively.

These bundles admit an atlas with constant transition maps and the
above flat connections have vanishing symbols with respect to this
atlas.
\eLm

\bpf
Let us consider the two inequivalent representations of
$\B Z_2$
on
$\B C$
\beq
\rho^+(1) = 1 \,,
\quad
\rho^+(-1) = 1
\ssep{and}
\rho^-(1) = 1 \,,
\quad
\rho^-(-1) = -1 \,.
\eeq

Then, the quotient of the trivial Hermitian line bundle 
$\tif Q\rot = S^3 \car \B C \to S^3$ 
with respect to the above actions
of $\B Z_2$ yields, respectively, the associated trivial and non
trivial Hermitian line bundles over 
$\f S\rot$ 
\beq 
\f Q^+\rot = S^3 \ucar{\rho^+} \B C \ssep{and} \f Q^-\rot = S^3
\ucar{\rho^-} \B C \,. 
\eeq

Moreover, the natural flat principal connection of the principal
bundle $S^3 \to SO(3)$ yields two flat Hermitian connections
$\chi^+\rot$ and $\chi^-\rot$ on $\f Q^+\rot$ and $\f Q^-\rot \,,$
respectively.\QED 
\epf

\bPr\label{trivial and non trivial rigid quantum bundles} 
The pullback with respect to the projection 
$\f E\rig \to \f S\rot$
yields a trivial and a non trivial Hermitian line bundle 
\beq 
\f Q^+\rig \to \f E\rig 
\ssep{and} 
\f Q^-\rig \to \f E\rig \,, 
\eeq
which are equipped with the pullback flat Hermitian connections
$\chi^+\rig$ 
and 
$\chi^-\rig \,,$ 
respectively.\END 
\ePr

\bTh
Let $\f E\rig$ be non degenerate. Then, the only inequivalent
quantum structures are of the type
$(\f Q^+\rig, \K Q\Upa^+\rig)$
and
$(\f Q\Upa^-\rig, \K Q\Upa^-\rig) \,,$
with
\beq
\K Q\Upa^+\rig = 
\chi\Upa^+\rig + \coi A\Upa^+\rig \ten \B I\Upa
\ssep{and} 
\K Q\Upa^-\rig = 
\chi\Upa^-\rig + \coi A\Upa^-\rig \ten
\B I\Upa \,,
\eeq
where
$\chi\Upa^+\rig \,, \chi\Upa^-\rig$
are
the pullbacks of 
$\chi^+\rig \,, \chi^-\rig \,,$ 
and 
$A\Upa^+\rig \,, A\Upa^-\rig$ 
are two global horizontal potentials for 
$\Ome \,.$
\eTh

\bpf
According to
Proposition \ref{existence and classification},
inequivalent quantum structures are in bijection with the set
\beq
(i^2)^{-1}([\Ome\rig]) \; \car \; H^{1}(\f E\rig, \Rn) \big/
H^{1}(\f E\rig, \B Z) =
\B Z_2 \car \{0\}\,.
\eeq

More precisely, the 1st factor paramet\-rises admissible quantum
bundles and the 2nd factor paramet\-rises quantum connections.\QED
\epf

In the following, we will specify the two possible trivial and non
trivial cases by the superscripts
$+$
or
$-$
only when it is required by the context.
%--------------------------------------------------------------------%
\mypar{Degenerate case.} 
\label{Degenerate caseII}
%--------------------------------------------------------------------%
Next, we analyse the degenerate case, following the same lines of the
non degenerate case.

\bPr
We have countably many equivalence classes of complex line bundles
with basis
$\f E\rig$
and just one equivalence class of quantum bundles.
Namely, this is the trivial one.
\ePr

\bpf
The 2nd cohomology group of
$\f E\rig$
is
\bat{3}
H^2(\f E\rig, \B Z) 
&\seq H^2(\f S\rig, \B Z) 
&&\seq H^2\big(S^2,
\B Z\big) 
&&\seq \B Z
\\
H^2(\f E\rig, \Rn)
&\seq
H^2(\f S\rig, \Rn)
&&\seq
H^2\big(S^2, \Rn\big)
&&\seq
\Rn \,.
\end{alignat*}

Then, according to
Proposition \ref{existence and classification},
the equivalence classes of complex vector bundles are in bijection
with
$H^2(\f E\rig, \B Z) \seq \B Z$
and the equivalence classes of quantum bundles are in bijection with
$(i^2)^{-1}([\Ome]) = (i^2)^{-1}(0) = \{0\} \,.$\QED
\epf

\bTh
Let $\f E\rig$ be degenerate. Then, the only quantum
structure is of the type 
$(\f Q\rig, \K Q\Upa\rig) \,,$ 
with
\beq
\K Q\Upa\rig = \chi\Upa\rig + \coi A\Upa\rig \ten \B I\Upa \,,
\eeq
where 
$\chi\Upa\rig$ 
is the pullback of 
$\chi\rig$ 
and
$A\Upa\rig$ 
is a global horizontal potential for 
$\Ome\rig \,.$
\eTh

\bpf
According to
Proposition \ref{existence and classification},
inequivalent quantum structures are in bijection with the set
\beq
(i^2)^{-1}([\Ome\rig]) \; \car \; H^{1}(\f E\rig, \Rn ) \big/
H^{1}(\f E\rig, \B Z ) =
\{0\} \car \{0\}\,.
\eeq

More precisely, the 1st factor paramet\-rises admissible quantum
bundles and the 2nd factor paramet\-rises quantum connections.\QED
\epf
%--------------------------------------------------------------------%
\mypar{Distinguished representatives.}
\label{Distinguished representatives}
%--------------------------------------------------------------------%
In both non degenerate and degenerate cases, the following facts hold.

\bPr\label{null of the observed potential}
Let us consider a global observer
$o : \f E\rig \to J_1\f E\rig \,.$

The form 
$- \C K\rig[o] + \C Q\rig[o]$ 
turns out to be a global horizontal potential for
$\Ome\Nat \,.$

Then, in the particular case when
$F = 0 \,,$
we can choose a representative of the quantum structure
$(\f Q\rig, \, \K Q\Upa\rig)$
in each equivalence class, such that
$A\rig[o] = 0 \,.$

Hence, the quantum differential turns out to be just the covariant
differential
$\nab[\chi\rig]$
associated with the flat connection(s)
$\chi\rig$
and the observed quantum Laplacian turns out to be just the
(spacelike) scaled Bochner Laplacian
$\Del[\chi\rig, g\rig]$
of the quantum bundle induced by the flat connection(s)
$\chi\rig$
and the (spacelike) metric
$g\rig \,.$\END
\ePr

\bPr\label{tensor product quantum bundle}
The Hermitian quantum bundle can be written, up to an equivalence, as
the fibred complex tensor product over
$\f T$
\beq
\f Q\rig = \f Q\cen \uten{\f T} \f Q\rot \,,
\eeq
where
$\f Q\cen \to \f E\cen$
is a Hermitian (trivial) quantum bundle over
$\f E\cen$
and
$\f Q\rot \to \f E\rot$
is a Hermitian quantum bundle over
$\f E\rot \,.$\END
\ePr

Accordingly, each quantum section
$\Psi\rig$
can be written as a finite sum of tensor products of the type
$\Psi\cen \ten \Psi\rot \,,$
which represent quantum states with decoupled center of mass and
rotational modes.
%--------------------------------------------------------------------%
\subsection{Quantum dynamics}
\label{Quantum dynamics}
%--------------------------------------------------------------------%
\bsm
Now, we apply the machinery of ``covariant quantum mechanics'' to
each one of the above three possible choices of quantum structures.

We will not repeat the whole procedure, but only sketch the main
differences between the one--body case and the rigid body case.
As one can expect, the most remarkable facts are due to the splitting
$\f E\rig = \f E\cen \car \f S\rot \,.$

The approach will be formally similar in the three cases, but the
equations will provide different results, as we will see in the
next section.
\esm

Let us consider any one of the three cases of quantum structures
discussed in the previous section.

Thus, let us consider the quantum bundle
$\f Q\rig \to \f E\rig$
and the phase quantum connection
$\K Q\Upa\rig \,.$

We have the splitting into coupled translational and rotational
components
\beq
A\Upa\rig = A\Upa\cen + A\Upa\rot \,,
\sep{with}
A\Upa\cen : J_1\f E\rig \to T^*\f E\cen \,,
\quad
A\Upa\rot : J_1\f E\rig \to T^*\f E\rot \,.
\eeq

The above splitting yields several other splittings.
In particular, we can write
\beq
\ob\Del\rig = \ob\Del\cen + \ob\Del\rot
\ssep{and}
\E S\rig\,_0 = \E S\cen\, _0 + \baE S\rot\,_0 \,,
\eeq
where
\bal
\ob\Del\cen \, \psi
&=
G\cen\,_0^{ij} \,
(\der_i - i \, A\cen\,_i) \, (\der_j - \coi A\cen\,_j) +
\fr{\der_i(G\cen\,_0^{ij} \, \rtd {g\cen})}
{\rtd {g\cen}} \, (\der_j - \coi A\cen\,_j)
\, \psi \,,
\\
\ob\Del\rot \, \psi
&=
G\rot\,_0^{\alp\bet} \,
(\der_\alp - \coi A\rot\,_\alp) \,
(\der_\bet - \coi A\rot\,_\bet) +
\fr{\der_\alp (G\rot\,_0^{\alp\bet} \, \rtd {g\rot})}{\rtd {g\rot}}
\,  (\der_\bet - \coi A\rot\,_\bet) \, \psi \,,
\\
\E S\cen\,_0 \, \psi
&=
(\der_0 - \coi A\cen \, _0 +
\tfr12 \fr{\der_0 \rtd {g\cen}}{\rtd {g\cen}}) \, \psi \,,
\\
\baE S\rot\,_0 \, \psi
&=
(- \coi A\rot
\tfr12 \fr{\der_0 \rtd {g\rot}}{\rtd {g\rot}} +
\tfr12 \ob\Del\cen\,_0 +
\tfr12 (\ob\Del\rot\,_0 + k\, \rho\rot\,_0) \big) \, \psi
\,.
\end{align*}

\myskip

The Lie algebra of special phase functions
$\spec(J_1\f E\rig, \, \Rn)$
has two remarkable subspaces, namely
$\spec(J_1\f E\cen, \, \Rn)$
and
$\spec(J_1\f E\rot, \, \Rn) \,.$
These subspaces turn out to be subalgebras in the case  when
$\C F\rig$
is decoupled with respect to
$\f E\cen$
and
$\f S\rot \,.$
In this case, we have ``translational" and
``rotational" observables.

If
$f\cen \in \spec(J_1\f E\cen, \, \Rn)$
and
$f\rot \in \spec(J_1\f E\rot, \, \Rn) \,,$
then we have the coordinate expressions
\bal
f\cen &=
f^0\cen \, \tfr12 \, G\cen\,^0_{ij} \, x^i_0 \, x^j_0 +
f^i\cen \, G\cen\,^0_{ij} \, x^j_0 + \br f\cen
\,,
\\
f\rot &=
f^0\rot \, \tfr12 \, G\rot\,^0_{\alp\bet} \, x^\alp_0 \, x^\bet_0 +
f^\alp\rot \, G\rot\,^0_{\alp\bet} \, x^\bet_0 + \br f\rot \,.
\end{align*}

The associated quantum operators are
\bal
&\wha f\cen \, \psi =
\big(\br f\cen - \coi \tfr12 \der_j f^j\cen
- \coi f^j\cen \, (\der_j - \coi  A\cen\,_j) -
\tfr12 \, f^0\cen \, \ob\Del\cen\,_0 \big) \,
\psi \,,
\\
&\wha f\rot \, \psi =
\big(\br f\rot - \coi \tfr12 \der_\alp f^\alp\rot
- \coi f^\alp\rot \, (\der_\alp - \coi A\rot\,_\alp) -
\tfr12 \, f^0\rot \, (\ob\Del\rot\,_0 + k \, \rho\rot\,_0)\big)
\, \psi \,.
\end{align*}

In particular, we have the following special phase functions
\bgt
x^0, \, x^i\cen \in \spec(J_1\f E\cen, \, \Rn) \,,
\qquad
x^0, \, x^\alp\rot \in \spec(J_1\f E\rot, \, \Rn) \,,
\\
\C P\cen\,_j \,, \|\C P\cen\|^2 \in \spec(J_1\f E\cen, \, \Rn) \,,
\quad
\C P\rot\,_\alp \,,
\|\C P\rot\|^2 \in \spec(J_1\f E\rot \,, \Rn) \,,
\\
\C H\cen\,_0 \in \spec(J_1\f E\cen, \, \Rn) \,,
\quad
\C H\rot\,_0 \in \spec(J_1\f E\rot, \, \Rn) \,.
\end{gather*}
and the associated quantum operators
\bgt
\wha{x^i\cen} \, \psi = x^i \, \psi \,,
\qquad
\wha{\C P\cen\,_j} \, \psi =
- \coi (\der_j + \tfr12 \fr{\der_j\rtd {g\cen}}{\rtd {g\cen}}) \,
\psi \,,
\\
\wha{\C H\cen\,_0} \, \psi =
\big(- \tfr12 {\ob\Del\cen}{}\,_0 - A\cen\,_0\big) \, \psi \,,
\\
\wha{x^\alp\rot} \, \psi = x^\alp\rot \, \psi \,,
\qquad
\wha{\C P\rot\,_\alp} \, \psi =
- \coi (\der_\alp + \tfr12 \fr{\der_\alp\rtd {g\rot}}{\rtd {g\rot}})
\, \psi \,,
\\
\wha{\C H\rot\,_0} \,\psi =
\Big(
- \tfr12 \ob\Del\rot\,_0 + k \, \rho\rot\,_0 - A\rot\,_0
\Big) \, \psi \,,
\end{gather*}
and
\bgt
\wha{\|\C P\cen\|^2_0} \, \psi =
\big(G^{ij}\cen\,_0 \, A\cen\,_i \, A\cen\,_j -
\coi \big(\der_h A^h\cen +
2 \, A^h\cen \, (\der_h - \coi A\cen\,_h)\big) -
\ob\Del\cen\,_0\big) \, \psi
\\
\wha{\|\C P\rot\|^2_0} \, \psi =
\Big(G^{\alp\bet}\rot\,_0 \, A\rot\,_\alp \, A\rot\,_\bet -
\coi \big(\der_\alp A^\alp\rot +
2 \, A^\alp\rot \, (\der_\alp - \coi A\rot\,_\alp)\big) -
\ob\Del\rot\,_0 - k \, \rho\rot\,_0 \Big) \, \psi \,.
\end{gather*}
%--------------------------------------------------------------------%
\newpage
\section{Rotational quantum spectra} 
\label{Rotational quantum spectra}
%--------------------------------------------------------------------%
\subsection{Angular momentum in the free case} 
\label{Angular momentum in the free case}
%--------------------------------------------------------------------%
\bsm
In this section, we analyse the implementation of angular
momentum for a rigid body in the framework of covariant quantum
mechanics. We start by recalling the relevant facts concerning
angular momentum in covariant classical mechanics. In this case it
is well known that angular momentum appears as a conserved
quantity of systems which are invariant under rotations. More
precisely, in these systems the angular momentum can be
interpreted as a momentum map for the action of the rotation group.
If we assume that this momentum map takes values in the special
functions then we associate with every element of the Lie algebra of
the rotation group a quantum operator and we get in this way a Lie
algebra representation whose Casimir is the square angular
momentum operator.
\esm

In the present paper we restrict ourselves to the case when
$F = 0 \,,$
although our results are valid in greater generality.
The reader is referred to \cite{SalVit00} for further details on
symmetries in covariant classical mechanics.

We consider the following group actions
\[
SO(\f S\pat, g\pat) \car (\B T \car \f S\rot) \to \B T \car \f S\rot : (A,
(\tau, r)) \mto (\tau, A(r)) \,.
\]

We would like to find the invariance of the dynamical structures with
respect to the above action. To this aim, we choose a global
potential 
$A\Upa \,.$
We observe that 
$A\Upa$ 
splits into the
sum
$A\Upa = A\Upa\cen + A\Upa\rot$
in an obvious way.

\bPr
The group
$SO(\f S\pat, g\pat)$
is a group of symmetries of the potential
$A\Upa\rot \,.$
Moreover, the momentum map induced by the action of
$SO(\f S\pat, g\pat)$
is just the total angular momentum with respect to the center of
mass.
\ePr

\bpf
In fact,
$A\Upa\rot$
reduces to the kinetic energy of particles with respect to the center
of mass. It is not difficult to prove that it is invariant with
respect to orthogonal transformations (see \cite{CurMil85}). We have
the momentum map
\beq
J : \F{so}(\f S\pat, g\pat) \to C^{\infty}(J_1(\f T\times \f S\rot)) :
\ome \mto J(\ome) \eqv \ome^* \com \C P\rot \,.
\eeq
Here,
$\ome^* : \f S\rot \to T\f S\rot : r \mto \ome(r) \,;$
moreover,
$J_1(\f T \car\f S\rot) = \f T \car \B T^* \ten T\f S\rot \,.$
It is easy to show that
$\ome^*\com \C P\rot(v) = G\rot(\ome(r),v) \,,$
where 
$v \in \B T^* \ten T\f S\rot \,.$
We have the coordinate expression
\beq
J(\ome) = (G\rot)^0_{\alp\bet} \, x^\bet_0 (\ome^*)^\alp
\eeq

The Hodge star isomorphism yields a natural Lie algebra
isomorphism
$\F{so}(\f S\pat, g\pat) \seq  \B L^{-1} \ten \f S\pat$
sending the Lie bracket of
$\F{so}(\f S\pat, g\pat)$
into the cross product. In this way, if
$\ome \in \F{so}(\f S\pat, g\pat)$
and
$\ba\ome \in \B L^{-1} \ten \f S\pat$
is the corresponding element, then we can equivalently write
\beq
J : \B L^{-1} \ten \f S\pat \to \Cin(J_1(\f T \car \f S\rot)) :
\ba\ome \mto J(\ba\ome) \eqv G\rel(r \cro v, \ome) \,.
\eeq
This proves the last part of the statement.\QED
\epf

The map $J$ takes values into the space of special functions since
$J(\ome)$
is a linear function of velocities for each
$\ome \in \F{so}(\f S\pat, g\pat) \,.$
Hence, it makes sense to consider the lift of
$J(\ome)$
to a quantum operator.

More precisely, by a composition of the momentum map with the lift
of quantum functions to quantum operators we obtain the following
representation of the Lie algebra
$\F{so}(\f S\pat, g\pat)$
\[
\ha J : \F{so}(\f S\pat, g\pat) \to \op (\ha{\f Q},\ha{\f Q}) :
\ome \mto \wha{J(\ome)}.
\]
If we consider a global observer
$o : \f E\rig \to J_1\f E\rig$
then according to \cite{JanMod05p1} we have
\[
\wha{J(\ome)} = \coi \left( X[J(\ome)] \con
\nab[o] + \tfr12 \Dive \, X[J(\ome)]\right) + J(\ome)[o]
\]
but
$J(\ome)[o] = 0$
and
$\Dive \, X[J(\ome)] = 0$
since
$G^0\rot$
is a left invariant metric and
$X[J(\ome)]$
is the fundamental vector field associated with
$\ome\in \F{so}(\f S\pat, g\pat) \,.$
Therefore
\[
\wha{J(\ome)} = \coi X[J(\ome)] \con \nab[o] \,.
\]

Let us consider a basis
$\{\ome_1,\ome_2,\ome_3\}$
of the Lie algebra
$\F{so}(\f S\pat, g\pat)$
which be orthonormal with respect to the metric
$\sig\rot$
(recall that
$\sig\rot$
is isometric to
$-\tfr12 k_3 \,,$
where
$k_3$
is the Killing metric of
$SO(3)$).

\bDf
The square angular momentum operator
${\ha J}^2$
is
$\h^2 \, C \,,$
where $C$ is the Casimir of the Lie algebra representation
$\ha J : \F{so}(\f S\pat, g\pat) \to \op (\haf Q,\haf Q) \,,$
thus
\beq
{\ha J}^2 =
\h^2 \, C =
\h^2 \, \left(
\wha{J(\ome_1)} \com \wha{J(\ome_1)} +
\wha{J(\ome_2)} \com \wha{J(\ome_2)} +
\wha{J(\ome_3)} \com \wha{J(\ome_3)}
\right) \,.\END
\eeq
\eDf

\bNt
The differential operator $C$ is exactly the pullback to
$\f Q$
of the Bochner Laplacian
$\Del[\chi\rot]$
of the line bundle
$\f Q\rot \to \f S\rot$
with respect to the connection
$\chi\rot$
of
$\f Q\rot$
and the Riemannian metric
$\sig\rot$
of
$\f S\rot \,.$\END
\eNt
%--------------------------------------------------------------------%
\subsection{Energy in the free case}
\label{Energy in the free case}
%--------------------------------------------------------------------%

\bsm
In this section, we assume the simplifying hypothesis that the
electromagnetic field vanishes.
In such a case, the Schr\"odinger equation splits into the two
decoupled Schr\"odinger equations for the center of mass and
rotations.
Clearly, the first one is trivial.
So, we concentrate our attention just on the rotational Schr\"odinger
equation.

We evaluate the spectra of rotational Hamiltonian for both non
degenerate (for trivial and non trivial quantum bundles) and
degenerate cases.
\esm

In this section, we assume
$F = 0 \,.$

Moreover, we shall refer to a global inertial observer
$o : \f E\rig \to J_1\f E\rig$
and to a representative of the quantum structure
$(\f Q\rig, \K Q\Upa\rig)$
in the unique equivalence class, such that
$A\rig[o] = 0 \,,$
according to
Proposition \ref{null of the observed potential}.

\myskip

Thus, let us consider the quantum bundle 
$\f Q\rot \to \f S\rot \,,$ 
which may be trivial or not, and the associated sectional
quantum bundle 
$\whaf Q\rot \to \f T \,.$

Let us consider the quantum Hamiltonian operator
\beq
\wha{\C H\rot\,_0} = - \tfr12 (\ob\Del\rot\,_0 + k \,\rho\rot\,_0) \,,
\eeq
where, according to our choices,
\beq
\ob\Del\rot\,_0 =
\Del[\chi\rot, G^0\rot] =
\Del[\chi\rot, \tfr{m}{\h_0} \, g\rot]
\eeq
turns out to be just the (unscaled) metric Laplacian associated with
the flat connection
$\chi$
and the Riemannian metric
$G^0\rot \,.$
We stress that
$\Del[G^0\rot]$
does not depend on the choice of an observer, as
$\f S\rot$
is spacelike, while
$\ob\Del\rot\,_0$
depends on the choice of the observer
$o \,,$
which yields
$A\rig[o] = 0 \,.$
Thus, the above equality holds just for that observer.

\bLm\label{Riemannian covering and Laplacian spectra}
\cite[pag.145.]{BerGauMaz71}
Let
$(\tif M, \, \ti g) \to (\f M, \, g)$
be a Riemannian covering.
Then, the eigenfunctions of the Laplacian
$\Del[g]$
are the projections on
$\f M$
of the projectable eigenfunctions of the Laplacian
$\Del[\ti g] \,.$
Moreover, we have
$\Spec\Del[g] \sub \Spec \Del[\ti g] \,.$\END
\eLm

\bLm\label{Riemannian covering and Bochner Laplacian spectra}
Let
$(\tif M, \, \ti g) \to (\f M, \, g)$
be a Riemannian covering.
Let
$\f Q \to \f M$
be a complex line bundle obtained as quotient of the trivial line
bundle
$\tif Q \byd \tif M \car \B C \,,$
with respect to the equivalence relation induced by the covering.
Moreover, let us suppose that the bundle
$\f Q \to \f M$
is equipped with a flat connection
$\chi$
obtained as quotient from the trivial flat connection
$\ti\chi$ of the bundle
$\tif Q \to \tif M \,.$
Let us consider the Bochner Laplacians
$\Del[\ti\chi, \ti g]$
and
$\Del[\chi, g]$
of
$\tif Q$ and $\f Q \,,$
respectively. Then, the eigensections of the Laplacian
$\Del[\chi, g]$
are the projections to sections of
$\f Q \to \f M$
of the
projectable eigensections of the Laplacian 
$\Del[\ti\chi, \ti g] \,.$ 
Moreover, the corresponding eigenvalues are the same.\END
\eLm

\bLm\label{spectrum of the sphere}
\cite[pag.159,160.]{BerGauMaz71} Let 
$(\f S^n, \, \ti g) \sub (\Rn^{n+1}, \, \E g)$ 
be the standard sphere. Then, we have 
$\Spec \Del[\ti g] = \{\lam_d = - d \, (d+n-1) \st d \geq 0\} \,.$
Moreover, the eigenspace 
$\tiC H_d$ 
associated with 
$\lam_d$
consists of the restrictions to 
$\f S^n$ 
of harmonic homogeneous polynomials of degree 
$d$ 
of 
$\Rn^{n+1} \,.$
We have 
$\dim \tiC H_d = \binom{n+d-2}{d} (\frac{2d+n-1}{n-1}) \,.$\END 
\eLm

By 
Lemma \ref{Riemannian covering and Bochner Laplacian spectra}
we can identify the Casimir operator 
$C \,,$ 
which acts on sections of the line bundle 
$\f Q\rot \to \f S_\mathrm{rot} \,,$ 
with an operator
$\tilde C$ 
acting on functions on 
$S^3 \,.$ 
One can prove (see
\cite[Lemma 7]{Tej01}) that 
$\tilde C=\frac{1}{4}\Del[\ti g]$
where  
$\Del[\ti g]$ 
is the Laplacian of the standard Riemannian metric of 
$S^3 \,.$ 
Thanks to Lemma \ref{spectrum of the sphere} we have

\bTh
The spectrum of 
$\ha J^2$ 
is
\beq
\Spec( \ha J^2) = \big\{ \h \, j(j+1) \big\} \,,
\eeq
where 
$j \in \B N$ 
in the trivial case and
$j \in \tfr12 \B N$ 
in the non trivial case.

The complex multiplicity of the eigenvalue 
$j(j+1)$ is $(2j+1)^2 \,.$

The eigensections with eigenvalue 
$j(j+1)$ 
are the projections to
$\f Q\rot \to \f S\rot$ 
of the restrictions to 
$S^3$ 
of homogeneous harmonic complex polynomials in 
$\Rn^4$ 
of degree 
$2j$
in the trivial case and of degree 
$2j+1$ 
in the non trivial case.\END
\eTh

\bTh\label{spectrum in spherical case}
{\em Spherical, non degenerate case.\/}
The spectrum of 
$\wha{\C H\rot\,_0}$ 
is 
\beq
\Spec(\wha{\C H\rot\,_0}) = \big\{ E_j = \fr{\h_0}{2 \, I} \, j(j+1)
+ k \, \fr{3 \, \h_0}{2 \, I}\big\}\,, 
\eeq 
where 
$j \in \B N$ 
in the trivial case and 
$j \in \tfr12 \B N$ 
in the non trivial case.

The complex multiplicity of the eigenvalue
$E_j$
is
$(2j+1)^2 \,.$

The eigensections of
$E_j$
are the projections to
$\f Q\rot \to \f S\rot$
of the restrictions to
$S^3$
of homogeneous harmonic complex polynomials in
$\Rn^4$
of degree
$2j$
in the trivial case and of degree
$2j+1$
in the non trivial case.
\eTh

\bpf
We restrict our attention to 
$- \tfr12 \ob\Del\rot\,_0 \,,$
since the contribution of the scalar curvature is obvious.

In virtue of Proposition \ref{metric isomorphisms in the non
degenerate case} and formula \eqref{proportionality of metrics},
we have an isometry 
$\f S\rot \to SO(3) \,,$ 
with respect to the metrics 
$G\rot\,_0 = \fr m{\h_0} \, g\rot = \fr m{\h_0} \, \fr Im \,\sig\rot$
and 
$- \tfr12 \fr I{\h_0} \, k_3 \,,$ 
respectively. Hence, the standard two--fold Riemannian covering 
$S^3 \to SO(3)$ 
yields a two--fold Riemannian covering 
$S^3 \to \f S\rot \,,$ 
with respect to the metrics 
$- \tfr12\fr I{\h_0} \, g_3$ 
and 
$G\rot\,_0 = \fr m{\h_0} g\rot \,,$ 
respectively, where 
$g_3$ 
is the metric induced on 
$S^3$ 
by the Killing metric of 
$SU(2)$ 
via the natural identification 
$S^3\simeq SU(2) \,.$

We recall that
$\f Q\rot$
can be obtained from the trivial bundle
$S^3 \car \B C \to S^3$
by a quotient (see
Lemma \ref{trivial and non trivial rotational quantum bundles}).

If $\ti g$ is the standard metric of the sphere then one has 
$- \tfr12\, g_3=4\,\ti g \,.$ 
Therefore, the Theorem follows from Lemma
\ref{Riemannian covering and Laplacian spectra}, Lemma
\ref{Riemannian covering and Bochner Laplacian spectra} and Lemma
\ref {spectrum of the sphere}, by taking into account that the
eigenspace 
$\tiC H_d$ 
is projectable on 
$\f Q\rot$ if $d$ 
is even or odd in the trivial case or in the non trivial case,
respectively.\QED
\epf

\bCr {\em Spherical, non degenerate case.\/}
The eigensections
with eigenvalue $E_j$ are eigensections of the square angular
momentum operator with eigenvalue 
$\h^2_0 \, j(j+1) \,.$
\eCr

\bLm
Let us consider a symmetric rigid body.
Let
$(X_1, X_2, X_3) \sub \f V\anl$
be an orthonormal basis, with respect to
$g \,,$
where
$X_1$
has the direction of the symmetry axis.

The corresponding basis (denoted by the same symbol) of $T\f
S\rot$ turns out to be left invariant and such that
$G^0\rot(X_i,X_i) = \fr{I_i}{\h_0} \,.$

Then, we obtain 
\beq 
\ob\Del\rot\,_0 = \Del[\chi\rot, G^0\rot] =
\frac{\h_0}{2I}\,C + \big(\fr{\h_0}{I_3} - \fr{\h_0}{I}\big)
\big(X_3 \con \nab[\chi\rot] \com X_3 \con \nab[\chi\rot]\big) \,,
\eeq 
where 
$X_3 \con \nab[\chi\rot]$ 
is regarded in a natural way as a differential operator acting on
sections of 
$\f Q\rot \,.$
\eLm

\bpf
We can write
\beq
\Del[\chi\rot, G^0\rot] =
\fr{\h}{I_1}\big(X_1\con\nabla[\chi\rot]\big)^2 +
\fr{\h}{I_2}\big(X_2\con\nabla[\chi\rot] \big)^2 +
\fr{\h}{I_3}\big(X_3\con\nabla[\chi\rot] \big)^2 \,.\QED
\eeq
\epf

We say that an eigenvalue depending on two parameters has {\em
arithmetical degeneracy\/} if it can be obtained from different
pairs of values of the parameters.

We note that complex polynomials in $\Rn^4$ can be regarded as
complex polynomials in the variables 
$(z_1, z_2, \ba z_1, \ba z_2)$ 
\cite[pag.169]{Tej01}.

\bTh {\em Symmetric non degenerate case.\/}
The spectrum of
$\wha{\C H\rot\,_0}$ is 
\beq 
\Spec(\wha{\C H\rot\,_0}) =
\Big\{E_{j,l} = 
\fr{\h_0}{2I} \, j(j+1) + 
\fr{\h_0}{2}\big(\fr{1}{I_3} -
\fr{1}{I}\big)\,l^2 + 
k \, \h_0 \, \big(\fr2{I} - \fr{I_3}{2 \,
I^2}\big) \Big\} \,, 
\eeq 
where 
$j \in \B N$ and 
$l \in \B Z$ 
in the trivial case and 
$2j \in \B N \setminus 2\B N$ and 
$2l \in \B Z \setminus 2\B Z$ 
in the non trivial case.

In case that there is no arithmetical degeneracy, the multiplicity
of the eigenvalue 
$E_{j,l}$ is $2(2j+1) \,.$

Eigensections of
$E_{j,l}$
are the projections to
$\f Q\rot \to \f S\rot$
of complex homogeneous harmonic polynomials in
$\Rn^4$
of degree
$p$
in
$z_i$
and degree
$q$
in
$\ba z_i \,,$
with
$p + q = 2 j \,,$
such that
$X_3$
has eigenvalue
$l$
on them.
\eTh

\bpf
The result can be obtained in the same way as Theorem
\ref{spectrum in spherical case}, by using the above Lemma and the
fact that the operators 
$\Del[\chi\rot, G^0\rot]$ 
and
$\big(X_1\con\nabla[\chi\rot]\big)^2$ 
commute \cite{Tej01}.

Of course, the eigenvalues of
$\big(X_1\con\nabla[\chi\rot]\big)^2$ 
are square integers and square half--integers on 
$S^3 \,.$\QED
\epf

\bCr {\em Symmetric non degenerate case.\/}
The eigensections with
eigenvalue $E_{j,l}$ are eigensections of the square angular
momentum operator with eigenvalue 
$\h^2_0 \, j(j+1) \,.$
\eCr

Arithmetical degeneracy can occur if 
$I_3/(I_3 - I) \in \B Q \,.$ 
In this  case, we could have 
$E_{j,l} = E_{j',l'}$ 
for some
$j \neq j'$ or $l \neq l' \,.$ 
See \cite{Tej01} for more details about the computation and the
multiplicity of eigenvalues and eigensections.

\bNt
Let us consider the asymmetric non degenerate case.

There is no general solution for the spectral problem, but just a
general method by which finding the solution in each case.
Namely, it is possible to restrict the Laplace operator to
$p+q+1$--dimensional subspaces
$H^{p,q}$
of harmonic complex polynomials of
$\Rn^4$
which are of degree
$p$
in
$z_i$
and degree
$q$
in
$\ba z_i \,,$
restricted to
$S^3 \,.$

The eigenvalue problem is solved by finding the root of the
characteristic polynomial, which is of degree $p+q+1 \,.$ Of
course, the  complexity of this problem increases with $p$ and 
$q \,.$ 
See \cite{Tej01} for more details.\END
\eNt

For the sake of completeness, we mention also the following more
standard result \cite{BerGauMaz71}, which follows directly from
Lemma \ref{spectrum of the sphere}.

\bTh {\em Degenerate case.\/}
The spectrum of 
$\wha{\C H\rot\,_0}$
is 
\beq 
\Spec(\wha{\C H\rot\,_0}) = 
\big\{E_j = \fr{\h_0}{2I} \, j(j+1) + k \, \h_0 \, \fr{2}{I}\big\}
\,, 
\eeq 
where 
$j \in \B N \,.$

The multiplicity of the eigenvalue
$E_j$
is
$(2j+1)^2 \,.$

Eigensections of
$E_j$
are the harmonic complex polynomials in
$\Rn^3$
restricted to
$S^2$
of degree
$2j \,.$\END
\eTh

In this case, the system is again invariant under rotations and
admits a momentum map which can be interpreted as the angular
momentum. Proceeding in a similar way as above we get

\bCr {\em Degenerate case.\/}
The eigensections with eigenvalue
$E_{j}$ are eigensections of the square angular momentum operator
with eigenvalue 
$\h^2_0\, j(j+1) \,.$
\eCr
%--------------------------------------------------------------------%
\subsection{Spectra with electromagnetic field}
\label{Spectra with electromagnetic field}
%--------------------------------------------------------------------%
\bsm
If the electromagnetic field does not vanish the computations of the
spectra might become quite hard.
However, specific problems can be faced.

Here, we sketch typical evaluations, with reference to the
literature, showing how they can be rephrased in our framework.
Indeed, our non trivial bundle structure opens a possible geometric
interpretation of the `two--valued' wavefunctions, which seems to
be closely related to spin.
\esm

\bEx ({\em Stark and Zeeman effects.\/})
The energy spectrum of a charged rigid body rotating in a constant
external electric, or magnetic field can be computed in our framework
along the lines of the previous section. The results fit the
computations in coordinates that can be found in the literature
\cite{HajOpa91,ChoSmi65,LanLif67,Mes59}, but provide also a
mathematical framework for the half-integer part of the spectrum
(which is usually discarded by invoking mathematical reasons).

For example, the energy eigenvalue equation in \cite{HajOpa91} can be
reproduced by means of the analysis of the electromagnetic field
acting on a rigid system performed in section 
\ref{Induced electromagnetic field}.  
In the case of trivial quantum bundle, the computation of the spectrum
can be rephrased word by word in our scheme. In the non trivial case,
one should use half--integer values of angular momentum and repeat
exactly the same computations.

In order to include in our scheme the anomalous Zeeman effect for a
rigid body with spin rotating in a constant magnetic field, one
should implement the spin in our covariant quantum mechanics
according to
\cite{CanJadMod95}.\END
\eEx

\bEx ({\em Magnetic monopole\/}).
Let us consider a rigid body
with a fixed point at which a monopole is located. In this case we
consider as electromagnetic pattern field the field generated by
the monopole. The main difference with the other examples of
electromagnetic fields considered before is that the cosymplectic
form
$\Ome$
defines a non-trivial cohomology class. Therefore
$\Ome$
does not admit global potentials. However, after lifting all the
structures to
$S^3$
(which is a fibration over
$\f S\rot$
in both the non degenerate and degenerate cases), the quantum
bundle becomes trivial and the computations are performed very
much in the same way as for the free rigid body. Another
difference is that the fixed point in the rigid body reduces the
degrees of freedom to the rotational part. Therefore, in this case
the analysis of the rotational part gives a complete description
of the rigid body.

One of us \cite{Tej01} gave an exact solution to the spectral
problem of a rigid rotator in a magnetic monopole field. We just
recall that a magnetic monopole is a magnetic field which is
proportional to the radial vector field in three--dimensional
space.

Here follows the spectrum of the energy operator in the case of a
symmetric rigid body:
\beq
\Spec (\wha{\C H\rot\,_0}) =
\left\{E_{j,l}
= \fr{\h_0}{2I} \, j(j+1) + 
\fr{\h_0}2 \bigl(\fr1{I_{1}} - \fr1I\bigr) \, l^2
- \nu_0 \, \fr{\|\vec q\|}{I_3} \, l + 
\fr{\nu^2_0}{\h_0} \, \fr{\|\vec q\|^2}{2I_3} + 
k \, \rho\rot\,_0\right\}
\,,
\eeq
where
$j\geq 0 \,, -j \leq l \leq j \,, j,l$
are integers, for the trivial quantum
bundle, or half--integers, for the non trivial quantum bundle,
$\nu$ is the magnetic charge of the monopole and
$\vec q$
is the center of charge of the  monopole \cite{Tej01}.\END
\eEx
%--------------------------------------------------------------------%

%--------------------------------------------------------------------%

\newpage
\begin{thebibliography}{12345678}
\fz
%--------------------------------------------------------------------%
\bi{Are70}
\au{R. Arens}
\tp{A quantum dynamical, relativistically invariant rigid body system}
\pu{Trans. A.M.S. {\bf 147}, (1970), 153--201}

%--------------------------------------------------------------------%
\bi{BarMikShi01}
\au{S. P. Baranovskii, V. V. Mikheev, I. V. Shirokov}
\tp{Quantum Hamiltonian systems on $K$-orbits: semiclassical
spectrum of the asymmetric top}
\pu{Teor. i Mat. Fiz. {\bf 129} n. 1 (2001), 1311--1319}

%--------------------------------------------------------------------%
\bi{BarBozMar92}
\au{A. O. Barut, M. Bo\v zi\'c, Z. Mari'c}
\tp{The Magnetic Top as a Model of Quantum Spin}
\pu{Ann. of Phys. {\bf 214} (1992) 53--83}

%--------------------------------------------------------------------%
\bi{BarDur91}
\au{A. O. Barut, I. H. Duru}
\tp{Path integral quantization of the magnetic top}
\pu{Phys. Lett. A {\bf 158} (1991), 441--444}

%--------------------------------------------------------------------%
\bi{BerGauMaz71}
\au{M. Berger, P. Gauduchon, E. Mazet}
\tb{Le Spectre d'une Vari\'et\'e Riemannienne}
\pu{Lect. Not. Math. 194, Springer, 1971}

%--------------------------------------------------------------------%
\bi{BotTu82}
\au{R. Bot, L.W. Tu}
\tb{Differential forms in algebraic topology}
\pu{GTM 82, Springer--Verlag 1982, Berlin}

%--------------------------------------------------------------------%
\bi{BozArs94}
\au{M. Bo\v zi\'c, D. Arsenovi\'c}
\tp{Quantum Magnetic Top}
\bk{Quantization and Infinite--Dimensional Systems}
\ed{J.--P. Antoine}
\pu{Plenum Press, New York 1994, 223--229}

%--------------------------------------------------------------------%
\bi{CanJadMod95}
\au{D. Canarutto, A. Jadczyk, M. Modugno}
\tp{Quantum mechanics of a spin particle in a curved spacetime with
absolute  time}
\pu{Rep. on Math. Phys., {\bf 36}, 1 (1995), 95--140}

%--------------------------------------------------------------------%
\bi{Cas31}
\au{H. Casimir}
\tb{Rotation of a Rigid Body in Quantum Mechanics}
\pu{Ph. D. Thesis, Wolters, Groningen 1931}

%--------------------------------------------------------------------%
\bi{ChoSmi65}
\au{J. H. Choi, D. W. Smith}
\tp{Lower Bounds to Energy Eigenvalues for the Stark Effect in a
Rigid Rotator}
\pu{J. Chem. Phys. {\bf 43}, 10 (1965), S189--S194}

%--------------------------------------------------------------------%
\bi{CohDiuLal77}
\au{C. Cohen--Tannoudji, B. Diu, F. Lalo\"e}
\tb{Quantum mechanics, vol. I and II}
\pu{Interscience, 1977}

%--------------------------------------------------------------------%
\bi{CurMil85}
\au{W. D. Curtis, F. R. Miller}
\tb{Differentiable Manifolds and Theoretical Physics}
\pu{Acad. Press, 1985}

%--------------------------------------------------------------------%
\bi{deLTuy96}
\au{M. de Leon, G. M. Tuynman}
\tp{A universal model for cosymplectic manifolds}
\pu{J. Geom. Phys. {\bf 20} (1996), 77--86}

%--------------------------------------------------------------------%
\bi{DoeMan98}
\au{H.--D. Doebner, H.--J. Mann}
\tp{Vector bundles over configuration spaces: topological potentials
and internal degrees of freedom}
\pu{J. Math. Phys.  {\bf 38} (1997), 3943--3952}

%--------------------------------------------------------------------%
\bi{DubNovFom79}
\au{B. Dubrovine, S. Novikov, A. Fomenko}
\tb{G\'eom\'etrie contemporaine, M\'ethodes et applications, Vol. 1,
2, 3}
\pu{Editions MIR, Moscou, 1979}

%--------------------------------------------------------------------%
\bi{DuvElhGotTuy90}
\au{C. Duval, J. Elhadad, M. J. Gotay, G. M. Tuynman}
\tp{Nonunimodularity and the quantization of the pseudo--rigid body}
\bk{Hamiltonian Systems, Transformation Groups and Spectral Transform
Methods}
\ed{J. Harnad and J. Marsden}
\pu{Publications du CRM, Montreal 1990}

%--------------------------------------------------------------------%
\bi{Fro82}
\au{A. Fr\"olicher}
\tp{Smooth structures}
\pu{LNM 962, Springer--Verlag, 1982, 69--81}

%--------------------------------------------------------------------%
\bi{Gar79}
\au{P. L. Garc\'\i a}
\tp{Cuantificacion geometrica}
\pu{Memorias de la R. Acad. de Ciencias de Madrid, {\bf XI}, Madrid,
1979}

%--------------------------------------------------------------------%
\bi{Got86}
\au{M. Gotay}
\tb{Constraints, reduction and quantization}
\pu{J. Math. Phys \textbf{27} (8) (1986), 2051--2066}

%--------------------------------------------------------------------%
\bi{Got98}
\au{M. Gotay}
\tp{Obstruction to quantization}
\bk{Proc. of the VII Conf. on Diff. Geom. and its Appl.}
\pu{Brno 1998}

%--------------------------------------------------------------------%
\bi{Got64}
\au{T. Goto}
\tp{Bohr--Sommerfeld's Quantum Conditions and Rigid Body Rotation}
\pu{Nuovo Cim. {\bf XXXI}, 2 (1964) 397--401}

%--------------------------------------------------------------------%
\bi{GuiSte82}
\au{V. Guillemin, S. Sternberg}
\tp{Geometric Quantization and Multiplicities of Group 
Representations}
\pu{Inventiones Mathematicae \textbf{67} (1982), 515--538}

%--------------------------------------------------------------------%
\bi{HajOpa91}
\au{J. Hajnal, G. Opat}
\tp{Stark effect for a rigid symmetric top molecule: exact solution}
\pu{J. Phys. B: At. Mol. Opt. Phys. {\bf 24} (1991), 2799--2805}

%--------------------------------------------------------------------%
\bi{HanReg74}
\au{A.J. Hanson, T. Regge}
\tp{The relativistic spherical top}
\pu{Ann. Physics, {\bf 87} (1974), 498-566}

%--------------------------------------------------------------------%
\bi{Her44}
\au{G. Herzberg}
\tb{Molecular spectra and molecular structure
I. Spectra of diatomic molecules, II. Infrared and Raman spectra of
polyatomic molecules}
\pu{Van Nostrand, New York, 1944}

%--------------------------------------------------------------------%
\bi{Hun77}
\au{W. Hunziker}
\tp{The Schr\"odinger eigenvalue problem for $N$-particle systems}
\pu{Acta Phys. Austr., Suppl. XVII (1977), 43--71}

%--------------------------------------------------------------------%
\bi{Iwa99}
\au{T. Iwai}
\tp{Classical and quantum mechanics of jointed rigid bodies with
vanishing total angular momentum}
\pu{J. Math. Phys. {\bf 40}, n. 5 (1999), 2381--2400}

%--------------------------------------------------------------------%
\bi{JadMod92}
\au{A. Jadczyk, M. Modugno}
\tp{An outline of a new geometric approach to Galilei general
relativistic quantum mechanics},
\bk{Differential geometric methods in theoretical physics}
\ed{C. N. Yang, M. L. Ge and X. W. Zhou}
\pu{World Scientific, Singapore, 1992, 543--556}

%--------------------------------------------------------------------%
\bi{JadMod94}
\au{A. Jadczyk, M. Modugno}
\tp{Galilei general relativistic quantum mechanics}
\pu{Report of Department of Applied Mathematics, University of
Florence, 1994, 1--215}
\wm

%--------------------------------------------------------------------%
\bi{JanMod02a}
\au{J. Jany\v{s}ka, M. Modugno}
\tp{Uniqueness Results by Covariance in Covariant Quantum
Mechanics}
\bk{Quantum Theory and Symmetries}
\ed{E. Kapu\'scik, A. Horzela}
\me{Proc. of the Second International Symposium,
July 18--21, 2001, Krak\'ow, Poland}
\pu{World Scientific, London, 2002, 404--411}

%--------------------------------------------------------------------%
\bi{JanMod02c}
\au{J. Jany\v{s}ka, M. Modugno}
\tp{Covariant Schr\"odinger operator}
\pu{Jour. Phys.: A, Math. Gen, {\bf 35}, (2002), 8407--8434}

%--------------------------------------------------------------------%
\bi{JanMod05p1}
\au{J. Jany\v{s}ka, M. Modugno}
\tb{Covariant Quantum Mechanics}
\pu{book in preparation, 2005}

%--------------------------------------------------------------------%
\bi{JanMod05p2}
\au{J. Jany\v{s}ka, M. Modugno}
\tp{Hermitian vector fields and special phase functions}
\pu{preprint, 2005}
\ar{math-ph/0507070}

%--------------------------------------------------------------------%
\bi{JanModSal02}
\au{J. Jany\v{s}ka, M. Modugno, D. Saller}
\tp{Covariant quantum mechanics and quantum symmetries}
\bk{Recent Developments in General Relativity}
\ed{R. Cianci, R. Collina, M. Francaviglia, P. Fr\'e}
\me{Genova 2000}
\pu{Springer--Verlag, Milano, 2002, 179--201}

%--------------------------------------------------------------------%
\bibitem{JanModVit05}
\textsc{J. Jany\v{s}ka, M. Modugno, R. Vitolo}:
Semi--vector spaces, preprint 2005.

%--------------------------------------------------------------------%
\bi{Joh80}
\au{G. John}
\tp{On geometric quantization of the rigid body}
\bk{Group theoretical methods in physics}
\ed{Wolf}
\pu{Lect. Not. Phys. 135,  1980}

%--------------------------------------------------------------------%
\bi{KliKon84}
\au{S. Klimek-Chudy, W. Kondracki}
\tp{On the $\theta$-theories and the multivalued wave functions}
\pu{Jour. Geom. Phys. {\bf 1} n.3 (1983), 1--12}

%--------------------------------------------------------------------%
\bi{KobNom69}
\au{S. Kobayashi, K. Nomizu}
\tb{Foundations of differential geometry I - II}
\pu{Interscience Publishers, John Wiley \& Sons, New York, 1969}

%--------------------------------------------------------------------%
\bi{Kom01}
\au{I. V. Komarov}
\tp{Remarks on Kovalevski's top}
\pu{J. Phys. A {\bf 34} (2001), 2111--2120}

%--------------------------------------------------------------------%
\bi{Kos70}
\au{B. Kostant}
\tp{Quantization and unitary repre\-sentations}
\pu{Lectures in Modern Analysis and Applications
III, Sprin\-ger--Verlag, {\bf 170} (1970), 87--207}

%--------------------------------------------------------------------%
\bi{LanLif67}
\au{L. Landau, E. Lifchitz}
\tb{M\'ecanique quantique, Th\'eorie non relativiste}
\pu{\'Editions MIR, Moscou, 1967}

%--------------------------------------------------------------------%
\bi{LitRei97}
\au{R. G. Littlejohn, M. Reinsch}
\tp{Gauge fields in the  separation of rotations and internal motions
in the $n$--body  problem}
\pu{Rev. Mod. Phys. {\bf 69}, 1 (1997), 213--275}

%--------------------------------------------------------------------%
\bi{LitMit01}
\au{R. G. Littlejohn, K. A. Mitchell}
\tp{Gauge theory of small vibrations in polyatomic molecules}
\bk{Proc. conf. in honour of the 60th birthday of J. Marsden}
\pu{408--428}

%--------------------------------------------------------------------%
\bi{Mar96}
\au{P. Maraner}
\tp{Monopole Gauge Fields and Quantum Potentials Induced by the
Geometry in Simple Dynamical Systems}
\pu{Ann. of Phys. {\bf 246} (1996), 325--346}

%--------------------------------------------------------------------%
\bi{MarRat94}
\au{J. Marsden, T. Ratiu}
\tb{Introduction to Mechanics and Symmetry}
\pu{Texts in Appl. Math. 17, Springer 1994}

%--------------------------------------------------------------------%
\bi{Mar03}
\au{A. Martens}
\tp{Quantization of an affinely-rigid body with constraints}
\pu{Rep. Math. Phys. {\bf 51}, n.2/3 (2003), 287--296}

%--------------------------------------------------------------------%
\bi{Mes59}
\au{A. Messiah}
\tb{Quantum Mechanics, vol. I and II}
\pu{Dunod, 1959}

%--------------------------------------------------------------------%
\bi{ModTejVit00}
\au{M. Modugno, C. Tejero Prieto, R. Vitolo}
\tp{Comparison between Geometric Quantisation and Covariant Quantum
Mechanics}
\bk{Lie Theory and Its Applications in Physics - Lie III}
\ed{H.-D. Doebner, V.K. Dobrev and J. Hilgert}
\me{Proc. of the Third International Workshop,
11 - 14 July 1999, Clausthal, Germany}
\pu{World Scientific, London, 2000, 155--175}
\ar{math-ph/0003029}

%--------------------------------------------------------------------%
\bi{ModVit05p}
\au{M. Modugno, R. Vitolo}
\tb{The geometry of Newton's law and rigid systems}
\pu{preprint, 2005}

%--------------------------------------------------------------------%
\bi{PanDra99}
\au{F. Pan, J. P. Draayer}
\tp{Algebraic Solutions for the Asymmetric Rotor}
\pu{Ann. of Phys. {\bf 275} (1999), 224--237}

%--------------------------------------------------------------------%
\bi{Pav04}
\au{M. Pav\v{s}i\v{c}}
\tp{Rigid particle and its spin revisited}
\ar{hep-th/0412324}

%--------------------------------------------------------------------%
\bi{Put93}
\au{M. Puta}
\tp{On the dynamics of the rigid body with a single rotor and an
internal torque}
\pu{Rep. Math. Phys. {\bf 32}, 3 (1993), 343--349}

%--------------------------------------------------------------------%
\bi{SalVit00}
\au{D. Saller, R. Vitolo}
\tp{Symmetries in covariant classical mechanics}
\pu{J. Math. Phys., {\bf 41}, 10, (2000), 6824--6842}
\ar{math-ph/0003027}

%--------------------------------------------------------------------%
\bi{SlaSla91}
\au{A. K. S\l awianowska, J. J. S\l awianowski}
\tp{Quantization of affinely rigid body in $n$ dimensions}
\pu{Rep. Math. Phys {\bf 29} (1991), 297--320}

%--------------------------------------------------------------------%
\bi{Sni80}
\au{J. Sniaticki}
\tb{Geometric quantization and quantum mechanics}
\pu{Springer--Verlag, New York, 1980}

%--------------------------------------------------------------------%
\bi{Sou70}
\au{J.--M. Souriau}
\tb{Structures des syst\`emes dynamiques}
\pu{Dunod, Paris 1970}

%--------------------------------------------------------------------%
\bi{TanIwa99}
\au{S. Tanimura, T. Iwai}
\tp{Reduction of Quantum Systems on Riemannian Manifolds with Symmetry
and Application to Molecular Mechanics}
\ar{math--ph/9907005}.

%--------------------------------------------------------------------%
\bi{Tej01}
\au{C. Tejero Prieto}
\tp{Quantization of a rigid body in a magnetic monopole}
\pu{Diff. Geom. and Appl., {\bf 14}, (2001), 157--179}

%--------------------------------------------------------------------%
\bi{Tot97}
\au{J. A. Toth}
\tp{Eigenfunction localization in the quantized rigid body}
\pu{J. Diff. Geom {\bf 43} (1997), 844--858}

%--------------------------------------------------------------------%
\bi{Tru97}
\au{M. Trunk}
\tp{Algebraic Constraint Quantization and the Pseudo--Rigid Body}
\ar{hep-th/9701112}.

%--------------------------------------------------------------------%
\bi{Vit99}
\au{R. Vitolo}
\tp{Quantum structures in Galilei general relativity}
\pu{Annales de l'Institut H. Poincar\'e, {\bf 70}, 1999}

%--------------------------------------------------------------------%
\bi{Woo92}
\au{N. Woodhouse}
\tb{Geometric quantization}
\pu{2nd Ed., Clarendon Press, Oxford, 1992}

%--------------------------------------------------------------------%
\end{thebibliography}
\end{document}